\documentclass[iop]{emulateapj}

\usepackage{graphicx,txfonts,amssymb,natbib,epsfig}

\shorttitle{The MUSIC of CLASH}
\shortauthors{Meneghetti et al.}

\begin{document}


\title{The MUSIC of CLASH: predictions on the concentration-mass relation}


\author{M. Meneghetti\altaffilmark{1,2,3}}
\author{E. Rasia\altaffilmark{4}}
\author{J. Vega\altaffilmark{5,33}}
\author{J. Merten\altaffilmark{2,9}}
\author{M. Postman\altaffilmark{6}}
\author{G. Yepes\altaffilmark{5}}
\author{F. Sembolini\altaffilmark{5}}
\author{M. Donahue\altaffilmark{12}}
\author{S. Ettori\altaffilmark{2,3}}
\author{K. Umetsu\altaffilmark{29}}
\author{I. Balestra\altaffilmark{32,19}}
\author{M. Bartelmann\altaffilmark{11}}
\author{N. Ben\'itez\altaffilmark{13}}
\author{A. Biviano\altaffilmark{32}}
\author{R. Bouwens\altaffilmark{21}}
\author{L. Bradley\altaffilmark{31}}
\author{T. Broadhurst\altaffilmark{22,23}}
\author{D. Coe\altaffilmark{6}}
\author{N. Czakon\altaffilmark{29}}
\author{M. De Petris\altaffilmark{8}}
\author{H. Ford\altaffilmark{31}}
\author{C. Giocoli\altaffilmark{28}}
\author{S. Gottl\"ober\altaffilmark{7}}
\author{C. Grillo\altaffilmark{16}}
\author{L. Infante\altaffilmark{24}}
\author{S. Jouvel\altaffilmark{14,15}}
\author{D. Kelson\altaffilmark{25}}
\author{A. Koekemoer\altaffilmark{9}}
\author{O. Lahav\altaffilmark{15}}
\author{D. Lemze\altaffilmark{31}}
\author{E. Medezinski\altaffilmark{31}}
\author{P. Melchior\altaffilmark{10}}
\author{A. Mercurio\altaffilmark{19}}
\author{A. Molino\altaffilmark{13}}
\author{L. Moscardini\altaffilmark{28}}
\author{A. Monna\altaffilmark{17,18}}
\author{J. Moustakas\altaffilmark{27}}
\author{L.A. Moustakas\altaffilmark{2}}
\author{M. Nonino\altaffilmark{32}}
\author{J. Rhodes\altaffilmark{1,9}}
\author{P. Rosati\altaffilmark{20}}
\author{J. Sayers\altaffilmark{9}}
\author{S. Seitz\altaffilmark{17}}
\author{W. Zheng\altaffilmark{6}}
\author{A. Zitrin\altaffilmark{9,30}}

\altaffiltext{1}{INAF, Osservatorio Astronomico di Bologna, via Ranzani 1, 40127 Bologna, Italy}
\altaffiltext{2}{Jet Propulsion Laboratory, California Institute of Technology, 4800 Oak Grove Drive, Pasadena, CA 91109, USA}
\altaffiltext{3}{INFN, Sezione di Bologna, Viale Berti Pichat 6/2, 40127 Bologna, Italy}
\altaffiltext{4}{Physics Dept., University of Michigan, 450 Church Ave, Ann Arbor, MI 48109, USA}
\altaffiltext{5}{Departamento de F'sica Te—rica, Universidad Aut—noma de Madrid, Cantoblanco, E-28049 Madrid, Spain}
\altaffiltext{6}{Space Telescope Science Institute, 3700 San Martin Drive, Baltimore, MD 21208, USA}
\altaffiltext{7}{Leibniz-Institut fŸr Astrophysik, An der Sternwarte 16, D-14482 Potsdam, Germany}
\altaffiltext{8}{Dipartimento di Fisica, Sapienza Universit‡ di Roma, Piazzale Aldo Moro 5, I-00185 Roma, Italy}
\altaffiltext{9}{California Institute of Technology, MC 249-17, Pasadena, CA 91125, USA}
\altaffiltext{10}{Center for Cosmology and Astro-Particle Physics and Department of Physics, The Ohio State University, Columbus, OH 43210, USA}
\altaffiltext{11}{Institut fur Theoretische Astrophysik, Universit\"{a}t Heidelberg, Zentrum f\"{u}r Astronomie, Philosophenweg 12, D-69120 Heidelberg, Germany}
\altaffiltext{12}{Department of Physics and Astronomy, Michigan State University, East Lansing, MI 48824, USA}
\altaffiltext{13}{Instituto de Astrof\'{\i}sica de Andaluc\'{\i}a (CSIC), E-18080 Granada, Spain}
\altaffiltext{14}{Institut de Ci\'{e}ncies de l'Espai (IEEC-CSIC), E-08193 Bellaterra (Barcelona), Spain}
\altaffiltext{15}{Department of Physics and Astronomy, University College London, London WC1E 6BT, UK}
\altaffiltext{16}{Dark Cosmology Centre, Niels Bohr Institute, University of Copenhagen, Juliane Maries Vej 30, DK-2100 Copenhagen, Denmark}
\altaffiltext{17}{Universit\"{a}ts-Sternwarte M\"{u}nchen, Scheinerstr. 1, D-81679 M\"{u}nchen, Germany}
\altaffiltext{18}{Max Planck Institute for Extraterrestrial Physics, Giessenbachstrasse, D-85748 Garching, Germany}
\altaffiltext{19}{INAF - Osservatorio Astronomico di Capodimonte, Via Moiariello 16, I-80131 Napoli, Italy}
\altaffiltext{20}{Dipartimento di Fisica e Scienze della Terra, Universit\`{a} degli Studi di Ferrara, Via Saragat 1, I-44122 Ferrara, Italy}
\altaffiltext{21}{Leiden Observatory, Leiden University, P. O. Box 9513, NL-2333 Leiden, The Netherlands}
\altaffiltext{22}{Department of Theoretical Physics and History of Science, University of the Basque Country UPV/EHU, P.O. Box 644, E-48080 Bilbao, Spain}
\altaffiltext{23}{Ikerbasque, Basque Foundation for Science, Alameda Urquijo, 36-5 Plaza Bizkaia, E-48011 Bilbao, Spain}
\altaffiltext{24}{Centro de Astro-Ingenier\'{\i}a, Departamento de Astronom\'{\i}a y Astrof\'{\i}sica, Pontificia Universidad Catolica de Chile, V. Mackenna 4860, Santiago, Chile}
\altaffiltext{25}{Observatories of the Carnegie Institution of Washington, Pasadena, CA 91101, USA}
\altaffiltext{26}{Institute for Computational Cosmology, Durham University, South Road, Durham DH1 3LE, UK}
\altaffiltext{27}{Department of Physics and Astronomy, Siena College, 515 Loudon Road, Loudonville, NY 12211, USA}
\altaffiltext{28}{Dipartimento di Fisica e Astronomia, Universit\`a di Bologna, Via Ranzani 2, 40127, Bologna}
\altaffiltext{29}{Institute of Astronomy and Astrophysics, Academia Sinica, P.O. Box 23-141, Taipei 10617, Taiwan}
\altaffiltext{30}{Hubble Fellow}
\altaffiltext{31}{Department of Physics and Astronomy, The Johns Hopkins University, 3400 North Charles Street, Baltimore, MD 21218, USA}
\altaffiltext{32}{INAF/Osservatorio Astronomico di Trieste, via G.B. Tiepolo 11, I-34143 Trieste, Italy}
\altaffiltext{33}{LERMA, CNRS UMR 8112, Observatoire de Paris, 61 Avenue de l'Observatoire, 75014 Paris, France}

\begin{abstract}
We present the results of a numerical study based on the analysis of the {\tt MUSIC-2} N-body/hydrodynamical simulations, aimed at estimating the expected concentration-mass relation for the CLASH cluster sample. We study nearly 1400  halos simulated at high spatial and mass resolution, which were projected along many lines-of-sight each. We study  the shape of both their density and surface-density profiles and fit them with a variety of radial functions, including the Navarro-Frenk-White, the generalised Navarro-Frenk-White, and the Einasto density profiles. We derive concentrations and masses from these fits and investigate their distributions as a function of redshift and halo relaxation. We use the X-ray image simulator X-MAS to produce simulated {\em Chandra} observations of the halos and we use them to identify objects resembling the X-ray morphologies and masses of the clusters in the CLASH X-ray selected sample. 
We also derive a concentration-mass relation for strong-lensing clusters. We find that the sample of simulated halos which resemble the X-ray morphology of the CLASH clusters is composed mainly by relaxed halos, but it also contains a significant fraction of un-relaxed systems. For such a heterogeneous sample we measure an average 2D concentration which is $\sim 11\%$ higher than found for the full sample of simulated halos. After accounting for projection and selection effects, the average NFW concentrations of CLASH clusters are expected to be intermediate between those predicted in 3D for relaxed and super-relaxed halos. Matching the simulations to the individual CLASH clusters on the basis of the X-ray morphology, we expect that the NFW concentrations recovered from the lensing analysis of the CLASH clusters are in the range $[3-6]$, with an average value of $3.87$ and a standard deviation of $0.61$. Simulated halos with X-ray morphologies similar to those of the CLASH clusters are affected by a modest orientation bias.
\end{abstract}


\section{Introduction}

Gravitational lensing is one the most powerful methods to investigate the distribution of matter (either dark or baryonic) in galaxy clusters. It is well known that this class of objects is particularly important in cosmology for several reasons. First, in a hierarchical model of structure formation, galaxy clusters are the latest bound structures to form in the universe. They are often captured in the middle of violent dynamical processes like mergers between smaller structures, allowing us to study in detail how structure formation proceeds. Second, each of them is a miniature universe, i.e. their composition closely reflects the matter composition of the universe at large. Last but not least, they trace the exponential tail of the structure mass function. Tiny variations of the cosmological parameters are reflected in dramatic changes of their mass function and of its evolution.

The lensing effects produced by galaxy clusters are sometimes spectacular. The light emitted by galaxies in the background of these objects interacts with the immense gravitational fields of these large cosmic structures and is deflected. Occasionally, if a background galaxy lays at small angular distance from the cluster center, the lensing effects are highly non linear, leading to the formation of {\em giant} arcs and multiple image systems. This regime is often called {\em strong lensing}. However, even at large angular distances the light feels the gravitational pull of the cluster. In this case, where the lensing distortion changes on scales much larger than the size of the sources, the shape of the distant galaxies is only weakly distorted. In this {\em weak lensing} regime, the lensing effects are described by means of an additional image ellipticity.   

Every cluster produces a weak lensing signal, while strong lensing events are rare and often observed only in the cores of the most massive clusters or in systems with enhanced shear fields. 
\cite{HE07.1} and \cite{2010A&A...519A..90M} illustrated with the help of numerical simulations how peculiar the population of strong-lensing clusters is. Clusters forming in the context of CDM  typically have oblate triaxial dark matter halos \citep{1988ApJ...327..507F,1991ApJ...378..496D,2011MNRAS.411..584M,2013SSRv..177..155L,2012ApJ...752..141L,2013MNRAS.431.1143D} and, among them, strong lenses tend to have their major axes preferentially oriented along the line-of-sight. Additionally, as described in \cite{TO04.1}, the cluster's ability to produce strong lensing features is boosted by dynamical events such as mergers or, more generally, by substructures orbiting around their host halo and occasionally crossing the cluster cores in projection \citep{2014ApJ...783...41B}. 

For these reasons, the selection of clusters based on their ability to produce strong lensing events is likely to generate a sample affected by biases. Since lensing is sensitive to the total mass projected on the lens plane, the halo structural  parameters inferred from the lensing analysis of clusters affected by an orientation bias will be biased as well. In particular, for clusters elongated along the line of sight, we expect to measure higher masses and concentrations \citep[see e.g.][]{2009ApJ...699.1038O,2009MNRAS.392..930O,HE07.1,2010A&A...519A..90M,2011ApJ...737...74G}, while the opposite is expected for clusters whose major axes are perpendicular to the line-of-sight.

To avoid this, a selection based on the cluster X-ray morphology is often advocated. 
The thermal X-ray emission by galaxy clusters originates in the Intra-Cluster-Medium (ICM), which is ionized gas heated to temperatures up to 100 keV emitting in the X-ray via thermal Bremsstrahlung \citep[e.g.][]{SA86.1}. In absence of processes inducing non-thermal pressure contributions, like e.g. perturbations induced by dynamical events like mergers or ICM turbulence, we do expect the ICM to be nearly in hydrostatic equilibrium with the cluster gravitational potential. As an indication for such equilibrium, or {\em relaxation}, the X-ray surface-brightness is expected to be symmetric and its iso-contours ``round" and concentric \citep[see e.g.][]{2013AstRv...8a..40R}. Following this philosophy, the CLASH cluster sample \citep{2012ApJS..199...25P} has been constructed  by selecting 20 massive clusters from X-ray based compilations of massive relaxed clusters. The  the relaxation state has been established on the basis of X-ray morphological estimators applied to {\em Chandra X-ray Observatory} images. 

Are these selection criteria really leading to a sample which is unbiased in terms of lensing masses and concentrations? \cite{2012MNRAS.426.1558G} have recently pointed out that, for randomly selected cluster samples, the concentration-mass relation derived from a two-dimensional lensing analysis is expected to have a lower amplitude compared to the intrinsic 3D concentration-mass relation. The reason is identified in the prolate triaxial shape of the cluster halos. Due to their prolateness, the probability of observing them elongated on the plane of the sky is higher than the probability of viewing them with their major axes pointing towards the observer \cite[some examples are shown in Fig.~10 of][]{2012MNRAS.425.2169G}. \cite{2013ApJ...776...39R} showed that selecting clusters according to their X-ray
luminosity not only increases the normalisation of the $c-M$ relation
with respect to a control sample but also returns a steeper slope.
This behaviour is explained by the fact that at fixed mass, the most
luminous clusters are also the most concentrated.  

In this paper, we aim at using a set of numerical  simulations of galaxy cluster sized halos, the {\tt MUSIC-2} simulation set, to better understand the expected properties of a sample of clusters having X-ray morphologies similar to  the CLASH sample. In particular, we wish to quantify the possible residual biases on the mass and on the concentration estimates due to the CLASH selection function.  This work has two companion papers\footnote[1]{To appear on arXiv/astro-ph the same day as this work.}: the strong-lensing and weak-shear study of CLASH clusters by Merten et al. (2014) and the weak-lensing and magnification study of CLASH clusters by Umetsu et al. (2014), where a comparison between our results and the observational analysis of the CLASH sample is presented.

The paper is structured as follows: in Section 2, we introduce the simulation set used in our analysis and we describe the methods used to measure the shape of density profiles in simulated halos; in Section 3, we introduce the CLASH cluster sample to which the simulations will be compared; in Section 4, we describe the morphological parameters used to construct a sample of X-ray selected clusters resembling the properties of the CLASH clusters; in Section 5, we describe the general properties of the halos in the simulated set and discuss their concentration-mass relation; in Section 6, we discuss the concentration-mass relation of strong lensing and X-ray selected halos; in Section 7, we use the X-ray morphology of the simulated clusters to predict the concentrations of the individual CLASH clusters. Finally, Section 8 contains our summary and conclusions.

\section{Simulations}

\subsection{The {\tt MUSIC-2} sample}
The {\tt MUSIC-2} sample \citep{2013MNRAS.429..323S,2013arXiv1309.5387S,2014MNRAS.439..588B} consists of  a mass limited sample of re-simulated halos selected from the MultiDark cosmological simulation.  This  simulation is dark-matter only and contains 2048$^3$ (almost 9 billion) particles in a (1$h^{-1}$Gpc)$^3$ cube. It was performed in 2010 using ART \citep{1997ApJS..111...73K} at the NASA Ames Research centre. All the data of this simulation are accessible from the online {\itshape MultiDark Database}\footnote[2]{www.MultiDark.org}. 
The run was done using the best-fitting cosmological parameters to WMPA7+BAO+SNI ($\Omega_M = 0.27$, $\Omega_b = 0.0469$, $\Omega_{\Lambda}= 0.73$, $\sigma_8 = 0.82$, $n = 0.95$, $h = 0.7$). This is the reference cosmological model used in the rest of the paper.

The halo sample was originally constructed by selecting all the objects in the simulation box, which are more massive than 10$^{15}\;h^{-1}M_\odot$ at redshift $z=0$. In total, 282 objects were found above this mass limit. All these massive clusters were re-simulated  both with and without radiative physics. The zooming technique described in \cite{2001ApJ...554..903K} was used to produce the initial conditions for the re-simulations. 
All particles within a sphere of 6 Mpc radius around the centre of each selected object at $z = 0$ were found in a low-resolution
version ($256^3$ particles) of the MultiDark volume. This set of particles
was then mapped back to the initial conditions to identify the Lagrangian
region corresponding to a $6\;h^{-1}$Mpc radius sphere centred
at the cluster centre of mass at $z = 0$. The initial conditions of the
original simulations were generated in a finer mesh of size $4096^3$. By doing so, the mass resolution of the re-simulated objects was improved by
a factor of 8 with respect to the original simulations. 
The parallel {\tt TREEPM+SPH GADGET} code \citep{SP05.1} was used
to run all the re-simulations. 

The {\tt MUSIC-2} sample exists in two flavours. In a first set of re-simulations baryons were added to the dark matter distributions extracted from the parent cosmological box and their physics was simulated via SPH techniques, without including radiative processes. A second set of re-simulations accounts  for the effects of radiative cooling, UV photoionization, star formation and supernova feedback, including the effects of strong winds from supernova.

In this paper, we focus our analysis on the non-radiative version of these simulations. Our choice is based on the fact that radiative simulations  without a proper description of energy feedback from AGNs generally produce un-realistically dense cores, due to the well known over-cooling problem \citep[see e.g.][]{2011ASL.....4..204B}. More recent simulations show that this problem is mitigated in simulations which simulate energy feedback from AGNs \citep{2010MNRAS.405.2161D,2011MNRAS.412.1965M,2014MNRAS.438..195P,2013ApJ...776...39R,2014MNRAS.438..195P}. This physical ingredient is not yet included in the {\tt MUSIC-2} sample. Moreover, our intention is to correlate the profile measurements with the strong-lensing efficiency of the simulated halos. 
\cite{2012MNRAS.427..533K}, comparing simulations with different treatments of baryonic processes, find that the addition of gas in non-radiative simulations does not change significantly the strong lensing predictions. However, gas cooling and star formation together significantly increase the number of expected giant arcs and the Einstein radii by a non-realistic amount, particularly for lower redshift clusters and lower source redshifts. 
Further inclusion of AGN feedback, however, reduces the predicted strong lensing efficiencies such that the lensing cross sections become closer to those obtained for simulations including only dark matter or non-radiative gas. The main requirements for this study are 1) a large number of highly resolved halos to accurately measure the profiles and determine the dependence of concentration on mass; 2) the presence of gas in the simulations in order to allow their X-ray analysis (see Sect. 4). For these reasons, we choose to use the non-radiative version of the {\tt MUSIC-2} sample. 

The mass resolution for these simulations corresponds  to m$_{\rm DM}$=9.01$\times$10$^8\;h^{-1}M_\odot$ and  to m$_{\rm SPH}$=1.9$\times$10$^8\;h^{-1}M_\odot$.  The gravitational softening was set to 6 $h^{-1}$ kpc for the SPH and dark matter particles in the high-resolution areas. 
Several low mass clusters have been found close to the large ones and
not overlapping with them. Thus, the total number of re-simulated
objects is considerably larger than originally identified in the parent cosmological box.  In total, there are 535  clusters
with M$>$10$^{14}\;h^{-1}M_\odot$  at $z$ = 0 and more than 2000 group-like objects
with masses in  the range 10$^{13}\;h^{-1}M_\odot<M_{vir}<$10$^{14}\;h^{-1}M_\odot$. In this study, we use a subsample of these halos, as explained below.

We have  stored snapshots for  15 different redshifts in the range 0 $\leq$ 
$z$ $\leq$ 9 for each re-simulated object. The snapshots which overlap with the redshifts of the CLASH clusters are at $z=0.250, 0.333, 0.429$ and $0.667$. 

The sample is complete above the mass thresholds given in Table~\ref{table:masses}. To extend our analysis towards smaller masses and being able to constrain the concentration-mass relation over a wider mass range, we analyse also halos with masses below the completeness limits. In particular, we use all halos with mass $M_{vir}>2\times10^{14}\;h^{-1}M_{\odot}$. Thus, we investigate a total number of 1419 halos, summing all halos at different redshifts. 
 
\begin{deluxetable}{cccc}[!t]
\tablecaption{Completeness mass limits and numbers of halos above the completeness mass limits in the {\tt MUSIC-2} sample.}
\tablehead{
\colhead{Redshift} & \colhead{Mass limit ($M_{vir}$)} & \colhead{Mass limit ($M_{200}$)} & \colhead{N. of halos} \\
& \colhead{[$h^{-1} M_\odot$]} &  \colhead{[$h^{-1} M_\odot$]} & 
}
\startdata
0.250 & $6.3\times10^{14}$ & $4.3\times10^{14}$ & 128 \\
0.333 & $6.4\times10^{14}$ & $5.1\times10^{14}$ & 97 \\
0.429 & $6.0\times10^{14}$ & $5.0\times10^{14}$ & 80 \\
0.667 & $3.9\times10^{14}$ & $4.0\times10^{14}$ & 89 \\
\enddata
\label{table:masses}
\end{deluxetable}

\subsection{Density profiles}
\label{sect:profiles}
\subsubsection{Generalities}
\cite{NA96.1} argued that the density profiles of numerically simulated dark-matter halos can be well fitted by an appropriate scaling of a ``universal" function over a wide range of masses. The function suggested to fit these profiles was later dubbed as the {\em Navarro-Frenk-White} density profile (NFW hereafter) and  is given by
\begin{equation}
	\rho(r) = \frac{\rho_s}{(r/r_s)(1+r/r_s)^2} \;,
	\label{eq:nfw}
\end{equation}
where $\rho_s$ and $r_s$ are the characteristic density and the scale radius of the halo. The profile is characterized by a logarithmic slope which is shallower than iso-thermal for $r \ll r_s$ and steeper than iso-thermal for $r \gg r_s$.  

Subsequent numerical studies \citep[see e.g. ][]{NA97.1} confirmed that the NFW function is appropriate to describe the profiles of {\em equilibrium} halos, i.e. of systems that are close to being in virial equilibrium, and is now widely used to characterize the shape of cluster-sized halos both in observations and in simulations. 

Along with the definition of the NFW density profile came that of the halo {\em concentration}, $c_{\Delta}=r_{\Delta}/r_s$, which is the ratio of the size of the halo, here defined as the radius enclosing a certain mean over-density $\Delta$ above the critical density of the universe, $\rho_{crit}(z)$.  The most appropriate value to describe the size of an equilibrium halo is its {\rm virial} radius, i.e. the radius within which the halo particles are gravitationally bound and settled into equilibrium orbits. In this case the virial over-density, $\Delta_{vir}$, is a function of cosmology and redshift \citep[][]{1998ApJ...495...80B,1997PThPh..97...49N}. To avoid this cosmological dependence, \cite{NA96.1} adopted the round number of $\Delta=200$, which is commonly used in the literature independently on the assumed cosmological model. In this paper, we will also define the size of the halos as $r_{200}$, which is the radius enclosing a mean density $\overline{\rho}=200\rho_{crit}(z)$.  \cite{2014arXiv1401.1216D} recently showed that rescaling clusters to this radius returns a self-similar inner density profile.

Despite the fact that the profiles of equilibrium halos are well described by the NFW function, a large fraction of halos formed in a cosmological box are far from having reached virial equilibrium \citep{2012arXiv1206.1049L, 2013arXiv1303.6158M}. \cite{2013MNRAS.tmp.2736B} discussed the dependence of this fraction on cosmology, finding that it is particularly sensitive to dark-energy. The reason is simply understood, being dark-energy affecting the formation and the growth of the cosmic structures. In the case of non-equilibrium halos, the NFW function gives a poorer description of the shape of the density profiles, and other functios involving a larger flexibility (i.e. additional free parameters) may result preferable. One example is the generalized NFW profile \citep[gNFW, ][]{1996MNRAS.278..488Z}, which is given by 
\begin{equation}
	\rho(r) = \frac{\rho_s}{(r/r_s)^\beta(1+r/r_s)^{3-\beta}} \;.
	\label{eq:gnfw}
\end{equation}
Compared to the NFW model, this profile is characterized by and additional parameter, namely the logarithmic inner slope $\beta$,
\begin{equation}
-\frac{{\rm d}\ln\rho}{{\rm d} \ln r} = \beta \;,
\label{eq:innslope_gnfw}
\end{equation}
which is radius independent.

A strong debate exists in the literature about the inner slope of the density profile of simulated halos \citep[see e.g.][]{MO98.1,NE11.1}. The advent of modern supercomputers allow us to push the mass and the spatial resolution of numerical simulations to unprecedented limits, and the new results indicate that there is a systematic deviation of the dark matter halo profiles from the form proposed by NFW \citep{2006AJ....132.2685M,2010MNRAS.402...21N}. The function that fits best such profiles is the {\em Einasto} function \citep{1989A&A...223...89E,2012A&A...540A..70R}, 
\begin{equation}
	\rho(r)=\rho_{-2}\exp\left\{-2n\left[\left(\frac{r}{r_{-2}}\right)^{1/n}-1\right]\right\} \;,
	\label{eq:einasto}
\end{equation}
which is characterized by a running logarithmic slope,
\begin{equation}
	-\frac{{\rm d}\ln\rho}{{\rm d} \ln r}\propto r^{1/n}\;,
		\label{eq:innslope_einasto}
\end{equation}
parametrized in terms of the index $n$. The amplitude of the profile is set by the density $\rho_{-2}$, which is the density at the radius $r_{-2}$, i.e. at the radius where the logarithmic slope of the density profile is -2. 

\subsubsection{The density profiles of the {\tt MUSIC-2} halos}
To describe the structural properties of the {\tt MUSIC-2} halos, we perform an analysis of their three-dimensional density profiles based on the  functional forms introduced in this Section. Such analysis is done by fitting the Eqs.~\ref{eq:nfw}, \ref{eq:gnfw} and \ref{eq:einasto} to the azimuthally averaged density profiles of the simulated halos. The code used to perform this analysis is the same used in another CLASH paper by Merten et al. (2014)\footnote[3]{Based on the open-source library {\tt Levmar}, {\tt http://users.ics.forth.gr/lourakis/levmar/}}. As is common practice in the literature \citep[e.g.][]{2013arXiv1312.0945L},  we minimize the function
\begin{equation}
	R^{2}_{3D}=\frac{1}{N_{\rm dof}}\sum_{i}[\log_{10}{\rho_i}-\log_{10}\rho(r_i,\vec{p})]^2\;,
		\label{eq:res3d}
\end{equation}   
where $\rho_i$ is the density measured in the $i-$th radial shell and $\vec{p}$ is the vector of parameters which are adjusted to derive the best-fitting function $\rho(r)$. In the case of the NFW profile, $\vec{p}=[\rho_s,r_s]$, while in the cases of the gNFW or Einasto profiles $\vec p=[\rho_s,r_s,\beta]$ and $\vec p=[\rho_s,r_s,n]$, respectively. $N_{\rm dof}$ is the number of degrees of freedom, i.e. the number of radii at which the profiles are evaluated minus the number of free parameters in the fit. 

When analyzing these three-dimensional density profiles, we perform the fit over the radial range [$\tilde{r}_{min}$,$\tilde{r}_{200}$], where $\tilde{r}_{min}=0.02R_{vir}$, and $\tilde{r}_{200}$ is the true $r_{200}$ of the halo.

A similar analysis is performed on the two-dimensional profiles, i.e. on the azimuthally averaged surface-density profile, $\Sigma_i$, corresponding to an arbitrary line-of-sight to the halo. The details of this analysis are discussed in the paper by Vega et al. (in prep.). In this case, the fitting functions are the projections of the functions in Eqs.~\ref{eq:nfw}, \ref{eq:gnfw} and \ref{eq:einasto}: 
\begin{equation}
	\Sigma(R)=2\int_0^{r_t}\rho(r=\sqrt{R^2+\xi^2}){\rm d}\xi \;,
\end{equation}
where $\xi$ indicates the spatial coordinate along the line-of-sight and $R$ is the projected radius. In the formula above, $r_t$ is a truncation radius, which is introduced to take into account that our halos are at the center of a cube with side-length $r_t=6 h^{-1}$Mpc comoving. The figure-of-merit function to be minimised in this case is
\begin{equation}
	R^{2}_{2D}=\frac{1}{N_{\rm dof}}\sum_{i}[\log_{10}{\Sigma_i}-\log_{10}\Sigma(R_i,\vec{p})]^2\;,
		\label{eq:res2d}
\end{equation}
In order to be consistent with the analysis done on the {\tt CLASH} clusters, we perform the two-dimensional fits over the radial range  $[20h^{-1}{\rm kpc},R_{vir}]$.  

For both the three- and the two-dimensional analyses, the best-fit parameters are used to compute the masses and the concentrations of the simulated halos. In the following, we identify the quantities estimated from these two analyses with the labels $3D$ and the $2D$, respectively. The best-fit masses are obviously obtained by integrating the best-fit density profiles,
\begin{equation}
	M=4\pi\int_{0}^{r_{200}}\rho(r,\vec{p}_{best})r^2{\rm d}r \;.
\end{equation}
The value of $r_{200}$ used here is derived by solving the equation
\begin{equation}
\frac{\int_{0}^{r_{200}}\rho(r,\vec{p}_{best})r^2{\rm d}r}{r_{200}^3}=\frac{200}{3}\times \rho_{\rm crit}(z) \;.
\end{equation}

Using its original definition (NFW), the concentration is the ratio between $r_{200}$ and the scale-radius, $r_s$. For the NFW profile, the scale radius corresponds to the radius where
\begin{equation}
-\frac{{\rm d}\ln\rho}{{\rm d} \ln r} = 2 \;,
\end{equation}
i.e. where the density profile has an isothermal slope. In the rest of the paper, we adopt the same definition also for the gNFW and for the {\em Einasto} profiles, 
\begin{equation}
c_{200} \equiv \frac{r_{200}}{r_{-2}} \;.
\end{equation}
Note that for the gNFW the following relation holds between $r_{-2}$, the scale radius $r_s$, and the inner slope $\beta$:
\begin{equation}
r_{-2} = (2-\beta)r_{s} \;.
\end{equation} 

\subsection{Lensing analysis}
The lensing analysis of the {\tt MUSIC-2} halos is described in details in Vega et al. (in prep.). For the purpose of this paper, we  use their estimates of the Einstein radii over a large number of projections per cluster. We also use their convergence profiles, properly rescaled into surface-density profiles, and their mass and concentrations based on the fits of the surface density profiles. The masses $M_{2D}$ and the concentrations $c_{2D}$ are the equivalent to the values derived from a comprehensive lensing analysis of real observations. Hence, we compare $M_{2D}$ and $c_{2D}$ to Merten et al. (2014) and Umetsu et al. (2014). 

For this work, we use our consolidated lensing simulation pipeline \citep[see e.g.][and references therein]{2010A&A...519A..90M}. Very shortly, the following steps are involved:
\begin{itemize}
\item we project the particles belonging to each individual halo along the desired line of sight on the {\em lens plane};
\item starting from the position of the virtual observer, we trace a bundle of light-rays through a regular grid of $2048\times2048$ covering a region of $1.5\times1.5\;h^{-1}$Mpc around the halo center on the lens plane;
\item using our code {\tt RayShoot} \citep{2010A&A...514A..93M}, we compute the deflection $\vec{\alpha}(\vec{x})$ at each light-ray position $\vec{x}$, accounting for the contributions from all particles on the lens plane;
\item the resulting deflection field is used to derive several relevant lensing quantities. In particular, we use the spatial derivatives of  $\vec{\alpha}(\vec{x})$ to construct the convergence, $\kappa(\vec{x})$, and the shear, $\vec{\gamma}=(\gamma_1,\gamma_2)$, maps. These are defined as:
\begin{eqnarray}
\kappa(\vec{x}) & = & \frac{1}{2}\left(\frac{\partial \alpha_1}{\partial x_1}+\frac{\partial \alpha_2}{\partial x_2}\right) \;, \\
\gamma_1(\vec{x}) & = & \frac{1}{2}\left(\frac{\partial \alpha_1}{\partial x_1}-\frac{\partial \alpha_2}{\partial x_2}\right) \;, \\
\gamma_2(\vec{x}) & = & \frac{\partial \alpha_1}{\partial x_2} = \frac{\partial \alpha_2}{\partial x_1} \;;
\end{eqnarray}
\item the lens critical lines are defined as the curves along which the determinant of the lensing Jacobian is zero \citep[e.g.][]{SC92.1}:
\begin{equation}
\det A = (1-\kappa-|\gamma|)(1-\kappa+|\gamma|) = 0 \;.
\end{equation}
In particular, the {\em tangential} critical line is defined by the condition $(1-\kappa-|\gamma|)=0$, whereas the {\em radial} critical line corresponds to the line along which $(1-\kappa+|\gamma|)=0$. In the following sections, we will often use the term {\em Einstein radius} to refer to the size of the tangential critical line. As discussed in \cite{2013SSRv..177...31M}, there are several possible definitions for the Einstein radius. In this paper, we adopt the {\em effective} Einstein radius definition \citep[see also][]{2012A&A...547A..66R}, 
\begin{equation}
\theta_E \equiv \frac{1}{d_{\rm L}}\sqrt{\frac{S}{\pi}}\;,
\label{eq:einsteinr}
\end{equation}
where $S$ is the area enclosed by the tangential critical line and $d_{\rm L}$ is the angular diameter distance to the lens plane.
\end{itemize}
All the lensing quantities are computed for a source redshift $z_{\rm s}=2$. 

In order to increase the statistics and to take into account possible projection effects, Vega et al. (in prep) study each halo under a large number of lines of sight. More precisely, they investigate 100 lines of sight for the halos above the mass completeness limits and 30 projections for those below the completeness limit. This implies that, for each halo, we have a catalog containing at least 30 measurements of the Einstein radius, projected mass, and projected concentration.

\subsection{X-ray analysis}

We build a mock X-ray catalogue by producing for each simulated cluster three {\it{Chandra}} events files corresponding to orthogonal projections aligned with the  Cartesian axes of the simulation. Due to excessive computational demand we cannot investigate all the lines of sight considered in Vega et al. (in prep.). The images are created by the X-ray MAp Simulator \citep[X-MAS,][]{GA04.1}, in which we utilize the ancillary response function and redistribution matrix function proper of ACIS-S3 detector \citep[for a complementary X-ray analysis of the {\tt MUSIC-2} sample, we refer the reader to][]{2014MNRAS.439..588B}.  The field-of-view (fov) covers (16 arcmin)$^2$. For the cosmology and redshifts analyzed, the fov size is equivalent to the following physical scales: $5.43\;h^{-1}$Mpc at $z=0.250$, $6.57\;h^{-1}$Mpc at $z=0.333$, $7.71\;h^{-1}$Mpc at $z=0.429$, and $9.57\;h^{-1}$Mpc at $z=0.667$.  The spectral emission is generated by adopting the MEKAL model in which we fix the redshift to the simulation's value and the metallicity to a constant value equal to 0.3 times the solar metallicity as tabulated by \cite{1989GeCoA..53..197A}. Finally, the contribution of the galactic absorption is introduced through a WABS model with $N_H=5 \times 10^{20}$ cm$^{-2}$ \citep[see e.g.][]{2009MNRAS.397.1876L}. The exposure time is set to 100 ks allowing a fair comparison with observations.

\section{The CLASH cluster sample}
The {\em Cluster Lensing and Supernova survey with Hubble} (CLASH) is a Multi-Cycle-Treasury program with the Hubble Space Telescope. During the HST cycles 18-20, 524 orbits were dedicated to observe 25 massive galaxy clusters. Among the goals of the program is to use the gravitational lensing properties of these objects to accurately constrain their mass distributions. In particular, one of the key objectives is to establish the degree of concentration of dark matter in the cluster cores, a key prediction of structure formation models. The survey is described in detail in \cite{2012ApJS..199...25P}. 

The targets of the CLASH programme were selected to minimise the lensing-based selection that favors systems with overly dense cores. Specifically, 20 CLASH clusters are solely X-ray selected. The X-ray-selected clusters are massive ($kT>5$ keV) and, in most cases, they appear to have a regular X-ray morphology. Five additional clusters are included for their lensing strength. These clusters have large Einstein radii ($\theta_E > 35"$) and were included to optimize the likelihood of finding highly magnified high-z ($z > 7$) galaxies. Using galaxy clusters as gravitational telescopes is another of the key objectives of CLASH, and the programme has given an extraordinary contribution to this field of research \citep{2012Natur.489..406Z,2012arXiv1211.2230B,2013arXiv1308.1692B,2013arXiv1307.5847S,2013ApJ...762...32C}.

For each CLASH  cluster, a large number of lensing constraints was collected, either from the HST, the Subaru \citep[e.g.][]{2013ApJ...777...43M} or ESO/WFI \citep{2013MNRAS.432.1455G} telescopes or from the CLASH-VLT spectroscopic program \citep{2013A&A...559L...9B}. 
Using these data of unprecedented quality, mass models for several CLASH targets  have been published over the last few years employing different methods of reconstruction \citep{2011ApJ...742..117Z,2012ApJ...755...56U,2012ApJ...749...97Z,2012ApJ...747L...9Z,2012ApJ...757...22C,2013ApJ...777...43M,2013ApJ...762L..30Z}. These techniques are based on strong, weak , or a combination of strong and weak lensing. 

In two companion papers, Merten et al. (2014) and Umetsu et al. (2014) focus on the analysis of the X-ray selected sub-sample of CLASH clusters. In Merten et al. (2014), a well tested reconstruction method \citep{2009A&A...500..681M,2010A&A...514A..93M,2011MNRAS.417..333M,2012NJPh...14e5018R} is used to combine weak-and-strong lensing constraints and derive the convergence maps of these clusters. Fitting the surface density profiles extracted from the maps, they  measure  masses and  concentrations of the CLASH clusters. As said, the X-ray selected CLASH clusters are ideal for this density profile analysis. In this paper, we analyse the {\tt MUSIC-2} halos sample with the intent of deriving theoretical expectations to compare to the results of the observational analysis of Merten et al. (2014) and Umetsu et al. (2014).  

\section{X-ray selection}

\subsection{X-ray morphological parameters}
\label{sect:xmorpho}     
One of the goals of this paper is to identify halos in the {\tt MUSIC-2} sample that closely resemble the X-ray properties of the clusters in the CLASH X-ray selected sample. Since this sample was selected to have a high degree of regularity in the X-ray morphology, we try to find equivalents in the simulations that mimic these X-ray characteristics.

We use five  parameters to measure the X-ray morphology in the soft-energy band ([0.5-2] keV)
images of our halos. These morphological parameters are evaluated within a physical radius $R_{\rm max}=500$ kpc following the same procedure adopted in the X-ray analysis of the CLASH clusters. The results of this analysis will be presented in details in a forthcoming paper by Donahue et al. (in prep.). The five parameters are:
\begin{enumerate} 
\item the centroid-shift, $w$, which assesses how much the centroid of the X-ray surface brightness moves when the aperture radius used to compute it decreases from $R_{max}$ to smaller values. It is defined as 
\begin{equation}
	w=\frac{1}{R_{\max}} \times \sqrt{\frac{\Sigma(\Delta_i-<\Delta>)^2}{N-1}}
	\label{eq:w}	
\end{equation}
where $N$ is the total number of apertures considered and $\Delta_i$ is the separation of the centroids computed within $R_{\rm max}$ and within the $i^{th}$ aperture;
\item The ellipticity, $e = 1-b/a$, where the axial ratio is equal to the ratio of the square root of the eigenvalues obtained by diagonalizing the inertia tensor of the X-ray surface brightness evaluated within $R_{max}$ \citep{1992ApJ...400..385B};
\item the X-ray-brightness concentration, which is the ratio between the
integral of the surface brightness $S$ within two apertures with radii $100$ kpc and $R_{max}$,
\begin{equation}
c_{X}=\frac{S(r<100 {\rm kpc})}{S(r<R_{max})}.
\label{eq:csb}
\end{equation}
\cite{2010ApJ...721L..82C}
\item and 5. the third and forth order power ratios, $P_3$ and $P_{4}$. These are the third and fourth order multipoles of the surface-brightness distribution within an aperture of radius $R_{\rm ap}=R_{max}$. 
The generic $m$-order power ratio ($m>0$) is  defined as $P_m/P_0$ with
\begin{equation}
P_m=\frac{1}{2m^2R^{2m}_{\rm ap}}(a_m^2 +b^2_m) \ \ \ \ {\rm and} \ \ \ \  P_0=a_0 \ln(R_{\rm ap})
\end{equation}
where $a_0$ is the total intensity within the aperture radius \cite{1996ApJ...458...27B}. The generic moments $a_m$ and $b_m$ are expressed in polar coordinates, $R'$ and $\phi'$, and given by
\begin{equation}
a_m(r)=\int_{R^{\prime} \leq R_{\rm ap}} S(x^{\prime}) R^{\prime} \cos(m \phi^{\prime}) d^2 x^{\prime},
\end{equation}
and
\begin{equation}
b_m(r)=\int_{R^{\prime} \leq R_{\rm ap}} S(x^{\prime}) R^{\prime} \sin(m \phi^{\prime}) d^2 x^{\prime}.
\end{equation}
\end{enumerate}
For a review about X-ray morphological parameters, we refer to \cite{2013AstRv...8a..40R}.

The five morphological parameters introduced above are combined to define a global degree of X-ray regularity. Such quantity is measured with respect to the mean of the simulated sample. Note that, with reference to the X-ray appearance, we use the term ``regular" to indicate halos with un-perturbed surface brightness distributions \citep{2012NJPh...14e5018R}. Very often, these halos are called ``relaxed". We do not use this term to differentiate from the classification  discussed in Sect.~\ref{sect:relaxation}. Regular halos have small centroid shift,  ellipticity, and power ratios. In addition they have large surface brightness concentration. Thus, we define the regularity parameter 
\begin{eqnarray}
M& = &\left(\frac{\log_{10}(w)-\langle\log_{10}(w)\rangle}{\sigma_{\log_{10}w}}\right)+\left(\frac{e-\langle e \rangle}{\sigma_{e}}\right) \nonumber + \\ 
& + & \left(\frac{\log_{10}(1/c_{X})-\langle\log_{10}(1/c_{X})\rangle}{\sigma_{\log_{10}1/c_{X}}}\right) + \nonumber \\
& + & \left(\frac{\log_{10}(P_3)-\langle\log_{10}(P_3)\rangle}{\sigma_{\log_{10}P_3}}\right) + \nonumber \\
& + & \left(\frac{\log_{10}(P_4)-\langle\log_{10}(P_4)\rangle}{\sigma_{\log_{10}P_4}}\right)\;,
\label{eq:morphopar}
\end{eqnarray}
similarly to the {\it M} parameter derived in \cite{2013AstRv...8a..40R}. In the formula above each morphological parameter, $p_i$, is compared to its mean over the simulated halos, $\langle p_i \rangle$, and rescaled by the standard deviation of its distribution, $\sigma_{p_i}$.

By plugging the parameters $p_{{\rm CLASH},i}$ measured on the X-ray images of the CLASH clusters into Eq.~\ref{eq:morphopar}, we  use the $M$ parameter  to quantify the regularity of  the CLASH clusters with respect to the simulations. The $M$ parameters of the CLASH X-ray selected clusters are listed in Table~\ref{tab:CLASH}. To construct a sample of CLASH-like clusters, we  select the simulated halos having  similar regularity parameter  as the observed clusters. 

For the purpose of matching simulated halos to each individual CLASH cluster, we  define the parameter $C_X$, which is defined as the distance, in parameter space, between each CLASH cluster and the simulated halos: 
\begin{eqnarray}
C_X & = & \sum_{i=1,5} \left(\frac{p_i-p_{{\rm CLASH},i}}{\sigma_{p_i}}\right)^2\;,
\label{eq:mpar}
\end{eqnarray}
where $p_i=[\log_{10}(w),e,-\log_{10}(c_X),\log_{10}P_3,\log_{10}P_4]$ are the morphological parameters discussed above and $\sigma_{p_i}$ their standard deviations.
As a result, the sample constructed via the  $M$ parameter has similar X-ray regularity to the CLASH sample. When we match halos using $C_X$ we identify only the simulated halos closest to each individual CLASH cluster in the morphological parameter space.  

\subsection{Non-radiative vs radiative simulations}
While our choice to use the non-radiative version of the {\tt MUSIC-2} halos is motivated by the need of avoiding biases caused by over-cooling, it is well known that hydrodynamical simulations like those employed here poorly describe several X-ray properties of real clusters \citep{2011ASL.....4..204B,2012arXiv1205.5556K}. For this reason,  we do not use gas temperatures or X-ray luminosities to match the CLASH clusters in our simulations. Our comparison is based solely on the X-ray morphology. 

To evaluate how this is influenced by gas physics, we  use the hydrodynamical simulations described in \cite{2010MNRAS.401.1670F}  and in \cite{2011MNRAS.418.2234B} \citep[see also][]{2012MNRAS.427..533K,2014MNRAS.438..195P} to evaluate how the morphological parameters used in this work change with more realistic physical treatments of the gas. These simulations, performed in the framework of a cosmological setting similar to that of the {\tt MUSIC-2} simulations, exist both in non-radiative and radiative versions. Contrary to the {\tt MUSIC-2} simulations, the effects of AGN feedback are also included in the radiative case. The sample is significantly smaller, though. 70 of these halos were recently processed with the X-MAS simulator, both in the non-radiative and radiative versions. We use this analysis to quantify the impact of radiative processes on the morphological parameters.

\vspace{0.2cm}
\begin{figure}[t]
 \includegraphics[width=1.0\hsize]{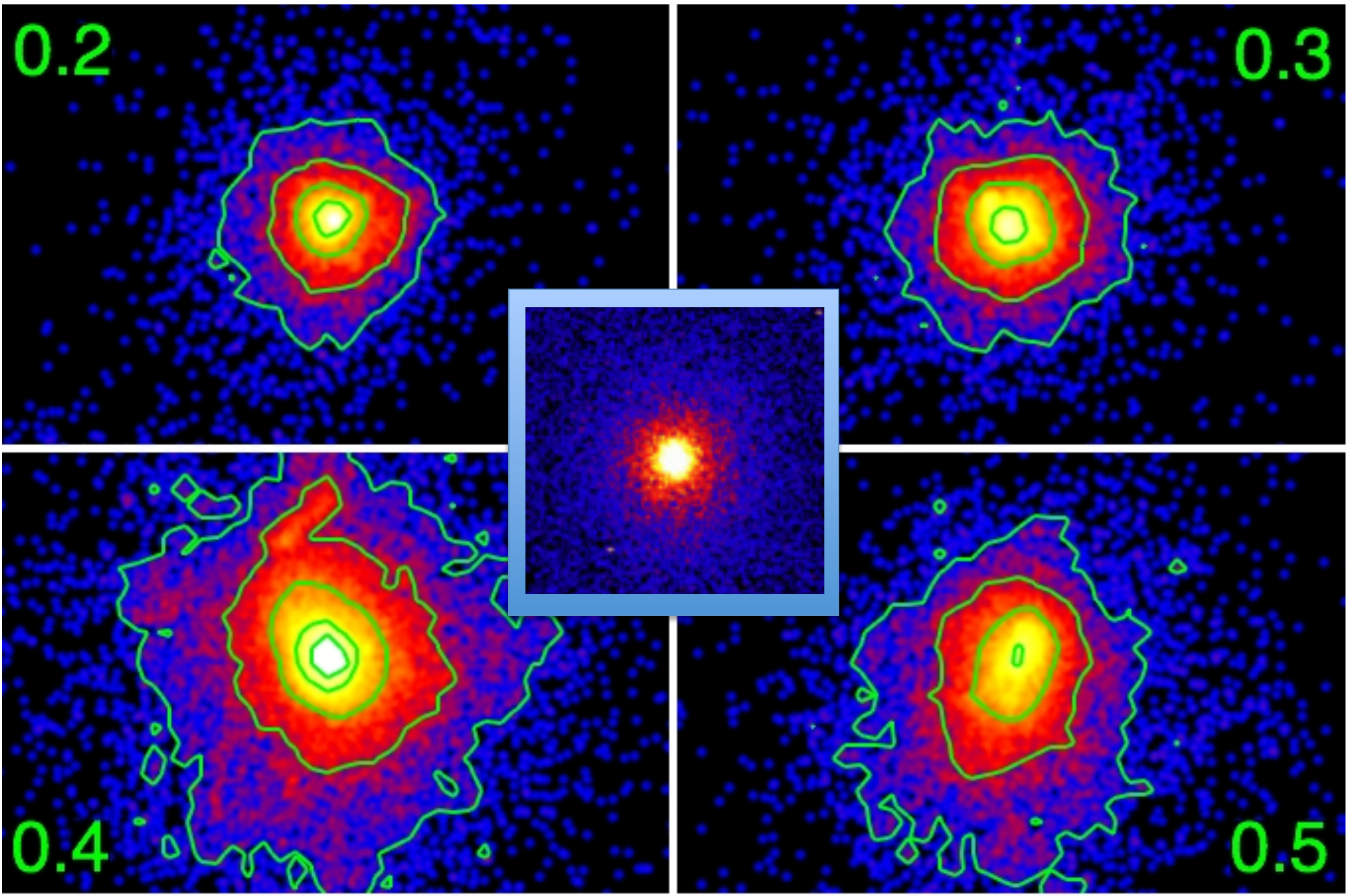}
 \caption{Examples of simulated clusters which match the CLASH cluster Abell 383 (shown in the small inset) with four increasing values of $C_X$.}
\label{fig:A383match}
\end{figure}

The distributions derived from the two simulated sets are consistent for all morphological parameters
computed within 500 kpc with the exception of the light concentration that is lower in the radiative simulation since part of the central
gas is turned into star and contributes less to the X-ray central emission. Applying the  selection method based on the parameter $C_X$ on the halos in these two datasets for a few CLASH clusters, we obtained identical matches. Therefore, we can assume that our X-ray selection method can safely be used on the non-radiative simulations.

\subsection{Example of regular cluster: Abell 383}
To illustrate how our selection based on the X-ray morphology performs, we discuss the case of Abell 383 \citep{2008MNRAS.383..879A}, which is the first cluster observed in the framework of the CLASH program. Abell 383 is a galaxy cluster at redshift $z=0.189$ \citep[see e.g.][]{2011ApJ...742..117Z}. In the X-ray, it exhibits a very regular morphology, with nearly circular surface brightness contours \cite[ellipticity $\sim 0.04$; ][]{2012ApJS..199...25P}. An X-ray image taken from the Archive of Chandra Cluster Entropy Profile Tables ({\tt ACCEPT}) is shown in the small inset at the center of Fig.~\ref{fig:A383match}. The image subtends $\sim 3.45'$.

The four largest panels of Fig~\ref{fig:A383match} show a sequence of simulated {\em Chandra}  observations of {\tt MUSIC-2} halos corresponding to increasing values of $C_X$, which are annotated on the images. The top left panel shows the X-ray morphology of the halo which best matches Abell 383 ($C_X=0.2$).  The X-ray morphology is indeed very similar to that of the observed cluster. As $C_X$ increases, the differences between the simulated and the true X-ray morphologies become more significant. On the basis of this and other visual inspections, we verified that $C_X\sim 0.4$ represents a good limit to select the halos ``similar" to the true  cluster. 




\subsection{Example of disturbed cluster: MACSJ1149}

\begin{figure}[t]
 \includegraphics[width=1.0\hsize]{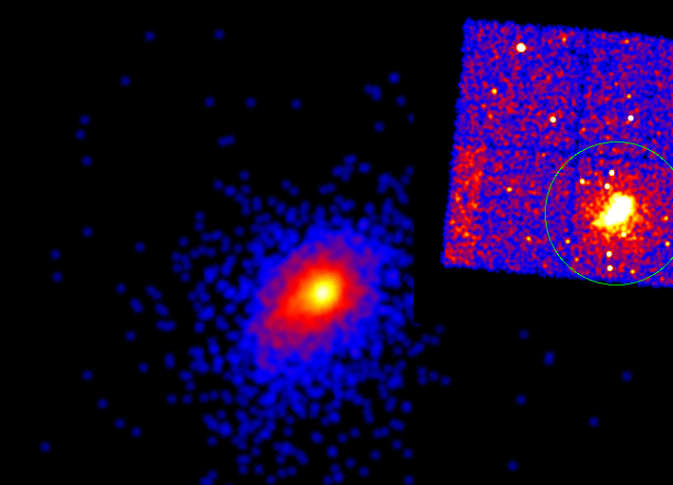}
 \caption{Best match to the morphologically disturbed cluster MACSJ1149. The real X-ray image of the cluster is shown in the small inset on the right.}
\label{fig:MACS1149match}
\end{figure}

Our selection  successfully identifies simulated halos that closely resemble also more perturbed clusters. For example, this is the case for MACSJ1149 \citep{2007ApJ...661L..33E}, which is one of the CLASH clusters identified as {\em high magnification clusters}, i.e. not included in the X-ray selected sample.  A comparison between the true X-ray morphology and that of a simulated halo with $C_X=0.18$ is shown in Fig.~\ref{fig:MACS1149match}, where we show the true {\em Chandra} image of the cluster in the smaller inset on the right. 

Clearly, the degree of asymmetry and of elongation of the surface brightness distribution in the simulated observation matches very closely that of MACSJ1149.

\section{Results}
In this section we discuss the results of our analyses on cluster mass profiles. First, we focus on the intrinsic properties of the whole sample, i.e. we do not apply any selection method to match the properties of the CLASH clusters. We compare to existing studies in the literature to verify the consistency of our and previous results. Then, we apply the selection based on the X-ray selection and perform a one-to-one comparison between the simulated halos and each CLASH cluster.

\subsection{Intrinsic properties of the {\tt MUSIC-2} halos}

\subsection{Relaxed and un-relaxed halos}
\label{sect:relaxation}
 In this Section we differentiate between relaxed and un-relaxed halos on the basis of a few criteria which are commonly used in the literature. Following the most restrictive approach proposed by  \cite{2007MNRAS.381.1450N}, we classify as strictly relaxed (or super-relaxed, as we dub them  later in the paper) those objects satisfying the following properties:
\begin{enumerate}
\item their centre of mass displacement, defined as the offset between the centre of mass (determined using all the particles within the virial radius) and the minimum of the potential, in units of the virial radius, is $s=(\vec r_{\rm cm}-\vec r_{\phi})/r_{\rm vir}<0.07$;
\item their virial ratio is $\eta=2T/|U|<1.35$, where $T$ is the kinetic energy and $U$ is the gravitational energy, computed using the particles within the virial radius;
\item their substructure mass fraction computed as the mass in resolved substructures within the virial radius, is $f_{\rm sub}<0.1$.
\end{enumerate}
Applying these selection criteria to the {\tt MUSIC-2} halos results into a fraction of relaxed halos of about $14.9\%$ at redshift $z=0.25$. The fraction is reduced to $11.7\%$ at redshift $0.333$ and it further drops to $10.4\%$ and to $8.9\%$ at redshifts $0.429$ and $0.667$, respectively. 

Other authors use less restrictive or alternative criteria to identify the relaxed systems \citep[e.g.][]{2011MNRAS.416.2388S,2011MNRAS.410..417S}. For example, \cite{2013ApJ...766...32B} only use the  centre of mass displacement.  In their paper, they report that the addition of the two other conditions on $\eta$ and $f_{\rm sub}$ does not affect significantly the selection. On the contrary, we find that using only the centre-of-mass displacement we end up with a significantly higher fraction of halos being classified as relaxed. This fraction amounts to $\sim 60\%$ at $z=0.250$ and decreases to $\sim 51\%$ at $z=0.667$.  Such fractions are compatible to those quoted by \cite{2013ApJ...766...32B} \citep[see also][]{2014MNRAS.439..588B}. \cite{2013arXiv1309.5387S} recently used the centre-of-mass displacement in combination with the virial ratio to identify relaxed systems in simulations. They report that the relation between $\eta$ and $s$ becomes flat for $s\lesssim 0.1$, thus indicating that $\eta$ does not impact severely on the selection of relaxed systems. 
For our sample, the combination of $s$ and $\eta$ yields to a fraction of relaxed halos corresponding to $47\%$ at $z=0.250$, which decreases to $29\%$ at $z=0.667$. 

In the following sections, we will study the properties of the {\tt MUSIC-2} halos dividing them into three sub-samples. First, we will consider all halos, regardless of the their relaxation state. Second, we will set the limit defined above on the centre-of-mass displacement to construct the sub-sample of {\em relaxed} halos. Third, we will further downsize the sample by using all three criteria described above, to identify the  {\em super-relaxed} halos.

\subsubsection{Density profiles}

\begin{figure*}[t]
 \includegraphics[width=0.49\hsize]{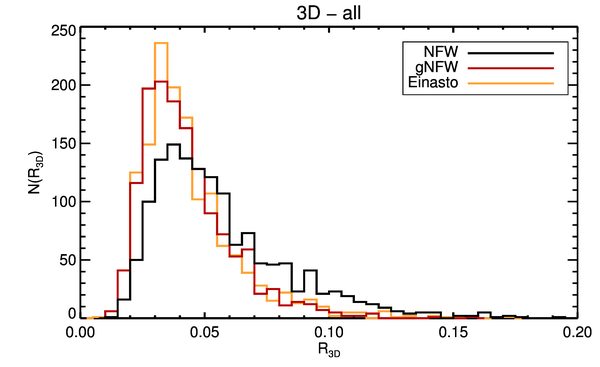}
 \includegraphics[width=0.49\hsize]{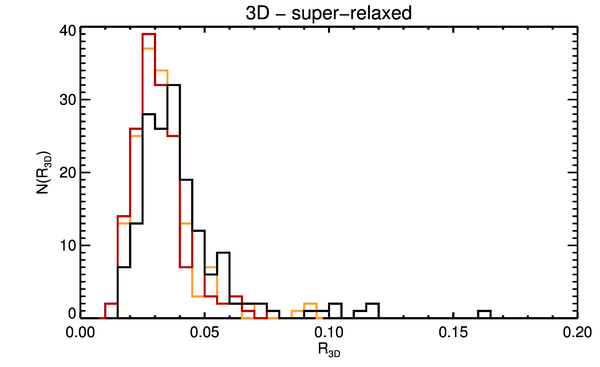}
 \includegraphics[width=0.49\hsize]{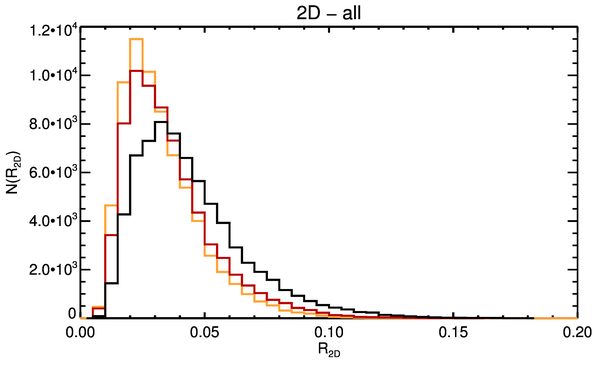}
 \hspace{0.2cm}
 \includegraphics[width=0.49\hsize]{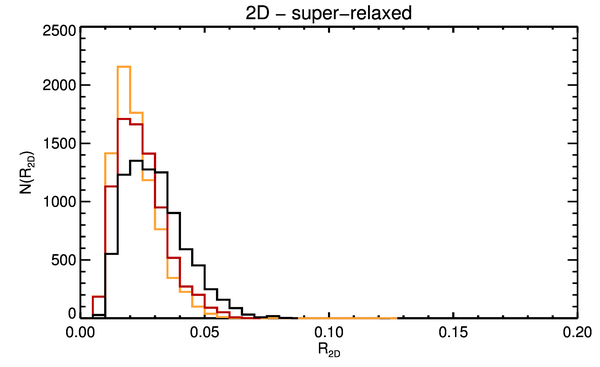}
 
 \caption{Distributions of the fit residuals. Results are shown for the fits of the density (upper panels ) and of the surface density profiles (bottom panels). The left and the right panels refer to the whole sample and to the subsample of super-relaxed halos, respectively. The black, red, and yellow histograms show the results for the NFW, gNFW, and Einasto models.}
\label{fig:histocomp}
\end{figure*}

As explained in Sect.~\ref{sect:profiles}, we fit the density profiles of the {\tt MUSIC-2} halos using the functions in Eqs.~\ref{eq:nfw}, \ref{eq:gnfw}, and \ref{eq:einasto}. In Fig.~\ref{fig:histocomp}, we show the results of the fitting procedure. We quantify the goodness of the fit by means of the residuals given in Eqs.~\ref{eq:res3d} and \ref{eq:res2d}.
 
The upper left panel shows the distributions of the fit residuals  for the entire {\tt MUSIC-2} sample. When all halos are considered, regardless of their relaxation state, the NFW profile is the worst fitting model, i.e. the one with the largest residuals \citep[see also][]{2013arXiv1303.6158M}. This is not surprising given that the NFW model has one free parameter less than the gNFW or the Einasto profiles. However, this result highlights the difficulty of fitting all profiles with a universal law. Since the gNFW and the Einasto functions generally provide better fits to the profiles, we may use the statistical distributions of their residuals to identify the halos deviating significantly from the NFW form. As it can be seen from Fig.~\ref{fig:histocomp}, such distributions are nearly log-normal, which suggests that halos having too large NFW residuals compared to the Einasto and the gNFW models may be identified via their deviation $\delta=\ln{R_{3D,{\rm NFW}}}-\langle\ln{R_{3D,x}}\rangle$, where $\langle\ln{{R}_{3D,x}}\rangle$ is the mean value of $\ln{R_{3D}}$ for either the Einasto or the gNFW model. Using this criterion, we find that about $40\%$ of the halos have NFW fits resulting in too large residuals compared to what typically found by fitting with more flexible profiles. 

This fraction drops to $\sim 19\%$ and  $\sim 6\%$ if only relaxed and super-relaxed halos are considered. The distributions of the fit residuals for the super-relaxed subsample are shown in the upper right panel of Fig.~\ref{fig:histocomp}. For these halos, the NFW model is only a slightly worse fit compared to the gNFW and Einasto models. 

\begin{figure}[t]
 \includegraphics[width=1.\hsize]{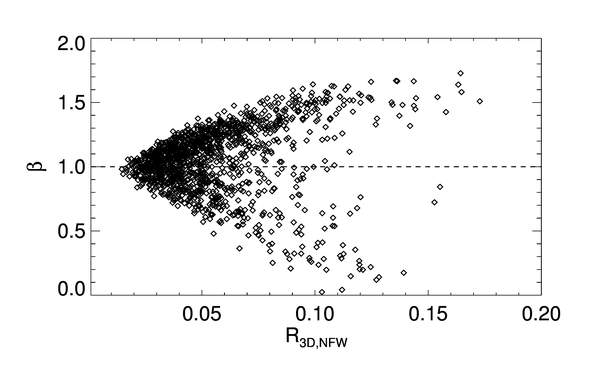} 
 \caption{Inner slopes, as they result from fitting the halo density profiles with gNFW models, vs the residuals of the NFW fits.}
\label{fig:slopes}
\end{figure}

\begin{figure}[t]
 \includegraphics[width=1.\hsize]{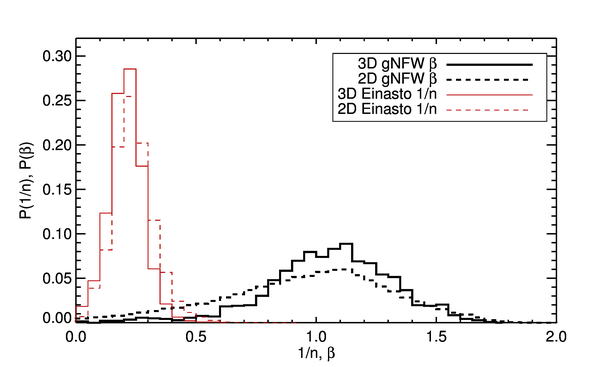} 
 \caption{Distributions of the inner slopes obtained from the gNFW fits ($\beta$) and of the Einasto index $1/n$ derived from the analysis of the density (solid histograms) and of the surface density profiles (dashed histograms) of the {\tt MUSIC-2} halos, as they result from fitting the halo density profiles with gNFW models, vs the residuals of the NFW fits.}
\label{fig:slopedistr}
\end{figure}

In Fig.~\ref{fig:slopes} we see that the profiles that most deviate from the NFW form have inner slopes $\beta$ (resulting from the gNFW fits) which significantly differ from unity: their profiles are steeper or shallower than the NFW model. There is a slight indication for preferring a steep  over a shallow slope (see also Fig.~\ref{fig:slopedistr}). Indeed, the mean value of the inner slope $\beta$, measured for the whole sample, is $\langle \beta \rangle=1.03\pm 0.31$.  We also find that the goodness of the gNFW fit is not correlated with the inner slope $\beta$, i.e. shallow or steep inner slopes are not systematically the result of a bad gNFW fit.   

When fitting the surface-density profiles, we find again that the NFW model is generally the worst fitting function among the three models employed in this work. This is shown in the bottom left panel of Fig.~\ref{fig:histocomp}. Again, we find that restricting the analysis to the relaxed halos  reduces the differences between the residual distributions of the NFW and gNFW or Einasto fits. However, from the results shown in the bottom right panel of Fig.~\ref{fig:histocomp}, it appears that  a fraction of  halos that are well fitted by NFW models in 3D are not NFW-like {\em in projection}. This result must be caused by the halo triaxiality and by the effects of substructures and additional matter along the line-of-sight. The work of Vega et al. (in prep.), from which the 2D analysis shown here is taken, investigates  the effects of triaxiality on shape of the surface density profiles of the CLASH clusters. We refer the reader to that paper for more details. We note that the halo surface-density profiles were derived by using all the particles in a cylinder centred on the halo and with depth $6 h^{-1}$Mpc. 

\begin{figure*}[t]
 \includegraphics[width=0.49\hsize]{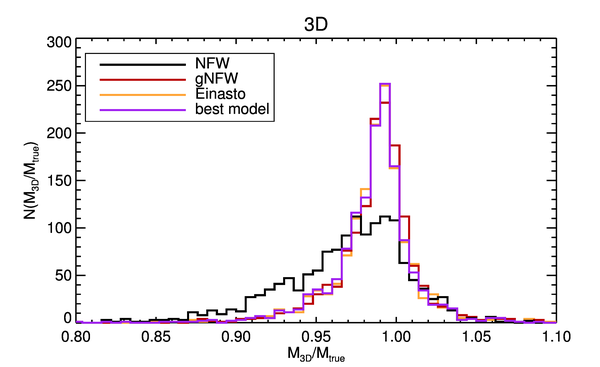}
 \includegraphics[width=0.49\hsize]{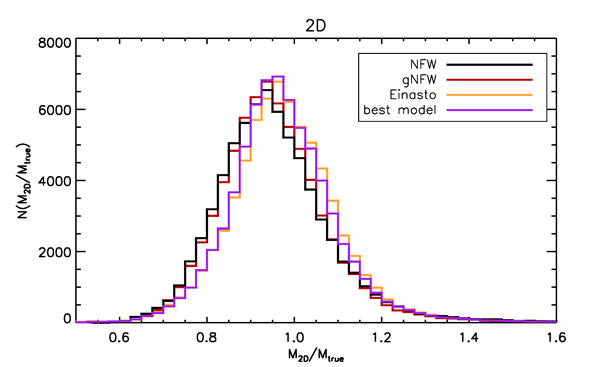} 
 \caption{Distributions of the ratios between 3D and true masses (left panel) and between 2D and true masses (right panel). The results are shown for the three fitting functions employed in this work: NFW (black), gNFW (red), and Einasto (yellow). We also show the distributions obtained for the mass estimates given by the model with the lowest residuals.}
\label{fig:masshistos}
\end{figure*}   

The distributions of the inner-slopes obtained from the gNFW fits of the surface density profiles are shown in Fig.~\ref{fig:slopedistr} (thick histograms). We find that a large number of halos have rather flat profiles in 2D. The mean value of $\beta$ is $\langle \beta \rangle = 0.89 \pm 0.47$. About $33\%$ ($15\%$) of the halo projections are fitted with $\beta \le 0.8$ ($\le 0.5$). The red histograms show the distributions of the Einasto indexes $1/n$. The indexes obtained from the fit of the density profiles are slightly smaller compared to what obtained from the fit of the surface density profiles. The smaller is $1/n$, the steeper is the inner profile. The mean values are $\langle 1/n \rangle = 0.21 \pm 0.07$ and $\langle 1/n \rangle = 0.24 \pm 0.09$ for the 3D and 2D distributions, respectively. Such Einasto slopes appear to be in excellent agreement with the recent results of \cite{2014arXiv1402.7073D}.
 
To summarise, the halos in the {\tt MUSIC-2} sample span a wide range of structural parameters. As expected, the density profiles can differ significantly from the NFW form and their shape can be better described with more flexible functions such as the Einasto or the gNFW models. When projecting the mass distributions, the scatter in the profile parameters and the deviation from the NFW model become even larger.

\subsection{Cluster masses}
\label{sect:masses}
Having determined the level of diversity between density and surface density profiles of the {\tt MUSIC-2} halos, we consider now how precisely the halo masses are derived form the profile fits. We consider both the cases of 2D and 3D masses, being the former the masses derived by de-projecting the best fit models of the surface density profiles under the assumption of spherical symmetry, and the latter those derived from the fits of the density profiles.
Note that, when measuring the 2D masses, we are not simulating any lensing analysis at this stage. In particular, we are not considering additional sources of systematics which may depend on the particular method to derive the mass from the weak and the strong lensing signals. Other works have shown that different methods of analysis may introduce systematic errors due, for example, to the presence of substructures inside and outside the clusters  \citep{2010A&A...514A..93M,2011ApJ...740...25B,2012NJPh...14e5018R} and to the Bright-Central-Galaxy \citep{2013arXiv1311.1205G}. Nevertheless, this exercise tells us important informations about the intrinsic limits of the mass measurements based on the analyses of azimuthally averaged density or surface density profiles.

We begin with the 3D masses. The distributions of the ratios between such masses and the true halo masses are shown in the left panel of Fig.~\ref{fig:masshistos}. 
The results are shown for the three fitting functions employed in this work (black, red, and orange histograms). We find that the masses recovered from the azimuthal fits of the density profiles are generally in good agreement with the true masses. The best agreement is obtained with the Einasto and gNFW profiles, with a slight preference for the first. These fits provide ratios around unity with r.m.s. 0.06  and 0.05, respectively. The masses estimated through the NFW fits are also in good agreement with the true masses. In this case the median (mean) ratio is 0.98 (0.97) and the distribution is twice as broad as in the two previous cases. 
The purple histogram is constructed by choosing, for each cluster, the mass estimate derived from the fitting function leading to the smallest residuals. In other words, we choose the most reliable mass estimate among those obtained with the three fitting functions. In most cases, the best model is the Einasto profile. Thus, the purple and the orange histograms are nearly coincident.

The histograms shown here refer to the whole halo sample, regardless of the relaxation state. As shown in the previous Sect., the density profiles of the relaxed halos are generally equally well fitted  by NFW, gNFW, or Einasto models. Indeed, restricting the analysis to these halos, we find smaller r.m.s for all three kinds of fit ($\lesssim 0.03$), with mean and median ratios very close to unity. Despite the fact that the fraction of relaxed halos varies with redshift, we find that the mean mass ratios and their scatter remain constant as a function of redshift.

Even when fitting the surface density profiles, the mass estimates ($M_{\rm 2D}$) deviate only slightly from the true masses. The 2D masses appear to be under-estimated by $\sim 5\%$ on average, with the NFW and the gNFW fits being slightly more biased than the Einasto fits. However, the scatter is much larger ($\sim 13-14\%$) than for the 3D masses.  The larger scatter is expected, given that the masses are derived under the assumption of spherical symmetry. Halos are generally triaxial and projection effects can easily cause the mass to be over- or under-estimated by a significant amount, depending on the halo orientation \citep[see e.g.][]{2010A&A...514A..93M}. As reported by \cite{2012MNRAS.426.1558G}, the halo prolateness may also cause a systematic under-estimation of the mass derived from the 2D analysis. Assuming the triaxial model of \cite{JI02.1} they estimate this bias to be of order  $\sim 10\%$.

As in the left panel, the purple histogram in the right panel of Fig.~\ref{fig:masshistos} shows  the distribution of the ratios between the best 2D mass estimate and  the true mass.  Again, the distribution is close to that obtained fitting with  the Einasto profile.

On the basis of this result, we conclude that we should expect a modest negative bias of $\sim 5\%$ on the mass estimates obtained  fitting the surface-density (or the convergence) profiles of galaxy clusters. This is due to the prolate shape of the halos, which are more frequently elongated on the plane of the sky than along the line-of-sight. The choice of the NFW or gNFW models to fit the halos tend to slightly increase the bias, while the opposite occurs with the Einasto profile.  

\begin{figure}[h]
 \includegraphics[width=1.\hsize]{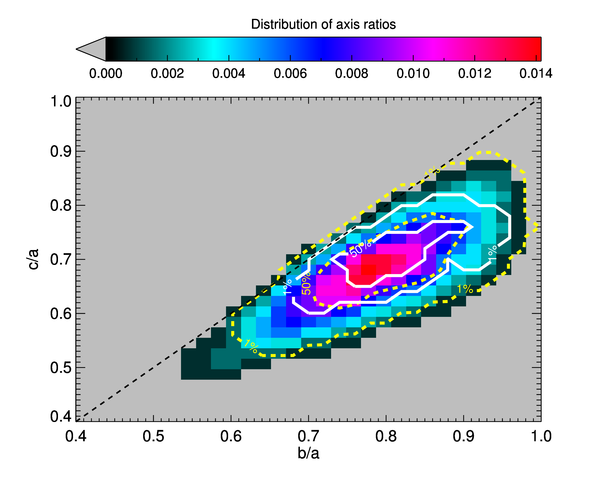}
 \caption{The color map shows the distribution of the axis ratios $b/a$ and $c/a$ of all the {\tt MUSIC-2} halos. The dashed and solid contours indicate the levels corresponding to $1\%$ and $50\%$ of the peaks of the distributions for the relaxed and super-relaxed halos.}
\label{fig:triaxiality}
\end{figure}

If we repeat this analysis on the samples of relaxed and super-relaxes halos, we find that the mass bias tends to become  smaller. In fact, the 2D masses deviate from the true masses by only $\sim1-2\%$ in these cases. If the bias is originated from halo triaxiality, this suggests that the most relaxed systems must  generally be more spherical. In Fig.~\ref{fig:triaxiality}, we show the distribution of the axis ratios $b/a$ and $c/a$ of all the {\tt MUSIC-2} halos (colour map).  Here $a$, $b$ and $c$ are the semi-axes of the inertial ellipsoid fitting the mass distribution of the halos with $a>b>c$. This fit is done using all particles  within the virial radius. It is clear from this plot that the relaxed  (yellow dashed contours) and the super-relaxed  systems (white contours)  generally have higher values of both $b/a$ and $c/a$. Thus, their shape is closer to spherical than that of non-relaxed halos, in agreement with \cite{2012ApJ...752..141L}.

\subsection{Concentration-mass relation}

The concentration-mass ($c-M-z$) relation is derived by means of nonlinear least-squares fitting using a Levenberg-Marquardt algorithm.  The fitting function we employ is 
\begin{equation}	
	c(M,z) = A\left(\frac{1.34}{1+z}\right)^B\left(\frac{M}{8\times10^{14}h^{-1}M_{\odot}}\right)^C \;, 
	\label{eq:cmfunction}
\end{equation}
which was also used by \cite{2008MNRAS.390L..64D} and by \cite{2013MNRAS.428.2921D}, although using a different pivot mass and redshift.  We perform this analysis for the three fitting models considered, and report the corresponding best fit parameters and errors in Table~\ref{tab:cmrelations}. The results are reported for the full sample as well as for the subsamples of relaxed and super-relaxed halos. We use Eq.~\ref{eq:cmfunction} to fit the $c-M-z$ relations derived from the analyses of the density profiles. 


\subsubsection{Comparison between fitting models}
\begin{figure}[t]
 \includegraphics[width=1\hsize]{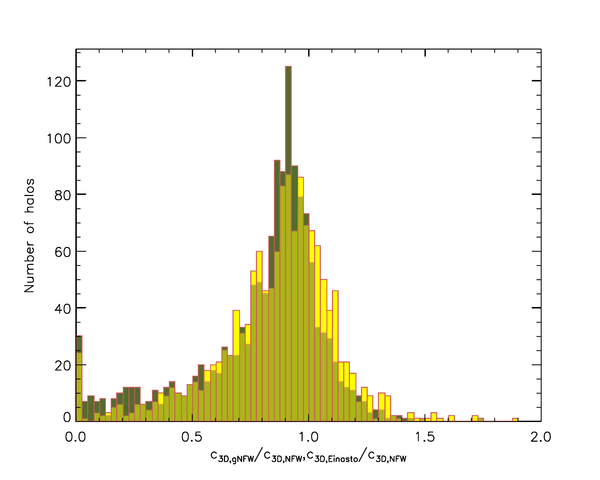}
  \includegraphics[width=1\hsize]{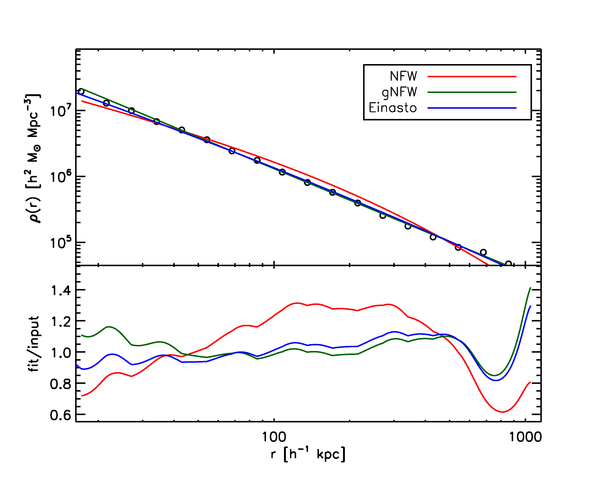}
 \caption{Upper panel: Distributions of concentration ratios $c_{\rm 3D,gNFW}/c_{\rm 3D,NFW}$ (yellow histogram) and $c_{\rm 3D,Einasto}/c_{\rm 3D,NFW}$ (green histogram). Bottom panel: Example of a density profile whose Einasto and gNFW concentrations are nearly zero. The halo profile is indicated by the open circles, while the best fit NFW, gNFW, and Einasto profiles are given by the red, green, and blue lines respectively. In the lower sub-panel we show the ratio between the best fit and the input profiles.}
 \label{fig:chisto}
\end{figure}

In the following, we consider the concentrations obtained from the NFW fit of the density profiles as reference, when making comparisons with the  concentrations derived from the gNFW and Einasto fits. The yellow and the green histograms in the upper panel of Fig.~\ref{fig:chisto} show the distributions of the ratios $c_{\rm 3D,gNFW}/c_{\rm 3D,NFW}$ and  $c_{\rm 3D,Einasto}/c_{\rm 3D,NFW}$ obtained from our analysis. In both cases, we find that the distributions peak at values around $\sim 0.9-0.95$, with the Einasto concentrations being generally smaller than the NFW ones. This result is in agreement with the recent findings of \cite{2014arXiv1402.7073D}, who also find that the Einasto concentrations are $\sim10-15\%$ smaller than the NFW concentrations on the mass scale of the {\tt MUSIC-2} halos. The halos with the smallest concentrations are of course the un-relaxed systems, for which we already pointed out that the NFW model is generally a bad fit. An example of such profiles is shown in the bottom panel of Fig.~\ref{fig:chisto}. In this case, the best fit NFW concentration is $c_{\rm 3D,NFW}=2.5$, while the gNFW and Einasto concentrations are $c_{\rm 3D,gNFW}=10^{-2}$ and $c_{\rm 3D,Einasto}=0.1$, respectively. Considering only the relaxed or the super-relaxed halos, the ratios between fitted and true concentrations are much closer to unity. For example, the mean ratios of $c_{\rm 3D,gNFW}/c_{\rm 3D,NFW}$ and  $c_{\rm 3D,Einasto}/c_{\rm 3D,NFW}$ for the super-relaxed systems are $1.0$ and $0.99$, respectively. We want to stress that the concentration of Einasto  profile being smaller than NFW does not imply necessarily that the halos are less concentrated. For the Einasto profile, the mass inside the scale radius also depends of the  $1/n$ parameter. An halo with the same mass ratio  between two radii as given by the NFW model, can be fitted with a smaller concentration and a larger $n$.

\begin{figure}[t]
\includegraphics[width=1\hsize]{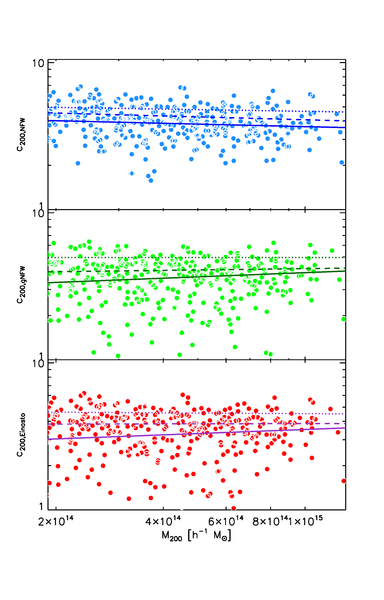}
\caption{Concentration-mass measurements at $z=0.250$. The results are shown for the full sample (filled circles). The upper, middle, and bottom panels refer to the NFW, gNFW, and Einasto fits, respectively. In each panel, we show the best fit $c-M-z$ relations for the full, relaxed, and super-relaxed samples (solid, dashed, and dotted lines, respectively).} 
\label{fig:cmfits}
\end{figure}

In Fig.~\ref{fig:cmfits}, we show the $c-M$ relations obtained from fitting the density profiles of the {\tt MUSIC-2} with the NFW, gNFW, and Einasto models (upper, middle, and bottom panels, respectively). The results are displayed for the halos at the lowest redshift investigated in this work ($z=0.250$). Each circle corresponds to a halo and the solid, dashed, and dotted lines indicate the best-fit $c-M-z$ relations for the full, relaxed, and super-relaxed sample.  At fixed mass, the distributions of NFW halos concentrations is reasonably well fitted by a log-normal distribution and have a standard deviation $\sigma_c \sim 0.25$, compatible with the findings of several previous works \citep[see e.g.][]{DO03.2}. The concentrations derived from the gNFW and the Einasto fits are characterised by a larger scatter. In all cases we find that the dependence of the concentration on mass is very shallow. For the NFW profile, $c \propto M^{-0.057\pm0.017}$ for the full sample.  Instead, for the gNFW and the Einasto profiles, the logarithmic slope of the $c-M$ relation is slightly positive. For the relaxed and super-relaxed halos, all the $c-M$ relations have logarithmic slopes which are negative or consistent with zero. As expected, we find that the more relaxed the halos are, the higher are their concentrations \citep{2009ApJ...707..354Z,2012MNRAS.422..185G}. This result holds regardless of the fitting function. At the lowest masses, the relative change in typical concentrations between the full and the relaxed (or super-relaxed) samples is larger for the gNFW and the Einasto fits. In fact, we find that a large fraction of small mass un-relaxed halos are fitted with lower concentrations using these two fitting models than with the NFW profile. These halos are responsible for the positive  logarithmic slope of the $c-M$ relation when fitting with the gNFW or Einasto  profiles.

As it can be seen from the $B$ parameters listed in Table~\ref{tab:cmrelations}, the normalisation of the 3D $c-M$ relation has an almost negligible redshift dependence for the full sample. For example, in the case of the NFW profile $c \propto (1+z)^{-0.29\pm0.08}$. For the gNFW and Einasto profiles, the redshift evolution is even shallower. We notice, however, that the dependence of the concentration on redshift appears to be stronger for the most relaxed systems.  In particular, for the super-relaxed halos we find $B \sim 0.52$ regardless of the fitting function.

\subsubsection{The NFW concentration-mass relation}
\begin{figure}[h]
\includegraphics[width=1\hsize]{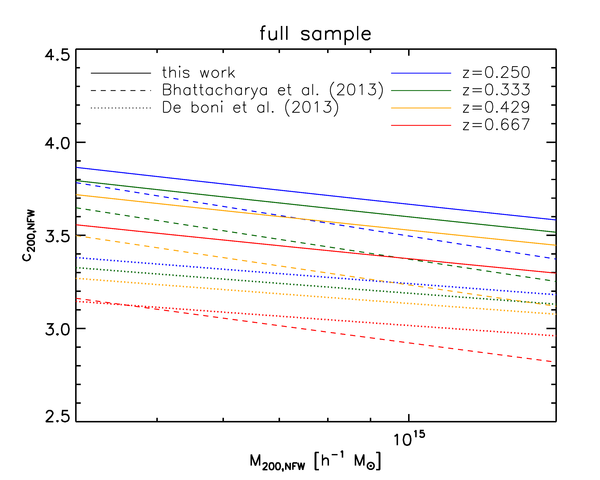}
\includegraphics[width=1\hsize]{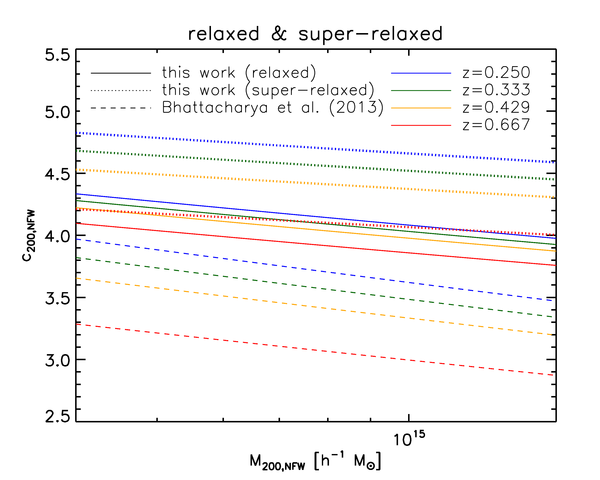}
 
\caption{Concentration-mass relation and its redshift evolution as obtained from fitting the halo density profiles with the NFW model. The results of this analysis are compared with the recent work of Bhattacharya et al. (2013) (dashed lines) and De Boni et al. (dotted-lines in the upper panel). The redshift evolution is illustrated by different colours. The upper and the bottom panels show the results for the whole sample and for the subsamples of relaxed and super-relaxed halos. Note that Bhattacharya et al. (2013) only distinguish between relaxed and un-relaxed halos.} 
\label{fig:3dvsBhat}
\end{figure}

There are several parameterisations of the $c-M$ relation in the literature, mostly derived from fitting simulated halos using NFW profiles. In the upper panel of Fig.~\ref{fig:3dvsBhat}, we show the NFW $c-M-z$ relation derived from the 3D analysis for the whole sample of {\tt MUSIC-2} halos (solid lines).  We use different colours to show how the relation evolves with redshift. 
We find a rather shallow dependence of the concentrations on  mass and redshift. Over the mass range $[4-12\times10^{14}h^{-1}M_{\odot}]$ the concentrations vary by less than $10\%$, decreasing as a function of mass as $M^{-0.058\pm0.017}$. The amplitude of the $c-M$ relation scales with redshifts as $(1+z)^{-0.29\pm0.08}$. Other authors find that the $c-M$ relation of massive halos is rather flat. For example, \cite{2009ApJ...707..354Z}, studying an ensemble of numerical simulations in the context of various cosmological models,  find that the concentration is strongly correlated with the age of the universe when the halo progenitor on the mass accretion history first reaches $4\%$ of its current mass. According to this correlation, they find that the concentration is nearly constant for halos with mass $M \gtrsim 10^{14} h^{-1} M_{\odot}$. They also predict a very shallow redshift evolution of the $c-M$ relation. In a recent work, \cite{2013MNRAS.428.2921D} also find concentrations that scale with mass and redshift similarly to our results. Their concentrations scale with mass and redshift as $M^{-0.07}$  and $(1+z)^{-0.26}$, respectively.   

The normalisation of our  $c-M-z$ relation is higher than found by some other authors like e.g. \cite{2013MNRAS.428.2921D} (dot-dashed lines in the upper panel of  Fig.~\ref{fig:3dvsBhat}) or \cite{2008MNRAS.390L..64D}. In these cases, the differences can be explained in terms of different cosmological settings. For example, \cite{2013MNRAS.428.2921D} analyse halos evolved in the framework of a WMAP3 cosmological model, and adopt a rather small normalisation of the matter power-spectrum, $\sigma_8=0.72$. If we consider other analyses in the literature in the context of WMAP7 normalised cosmologies, the agreement is much better. For example, the $c-M$ relation  which best fits our data at low redshift is in rather good agreement with the results of \cite{2013ApJ...766...32B} for non-relaxed halos. For comparison, their $c-M$ relation is over-plotted in the upper panel of  Fig.~\ref{fig:3dvsBhat} (dashed lines). At $z=0.250$, the concentrations we measure at a given mass are only $\lesssim 6\%$ higher than found by \cite{2013ApJ...766...32B}. However, their $c-M$ relation has a stronger redshift evolution. Between $z=0.250$ and $z=0.667$, their concentrations at a fixed mass decrease by $\sim 17\%$, while ours vary only by $\sim 10\%$.  

\begin{figure*}[t]
 \includegraphics[width=0.33\hsize]{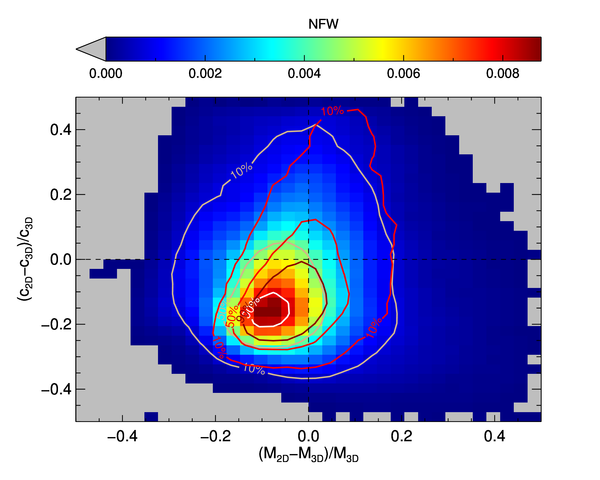}
  \includegraphics[width=0.33\hsize]{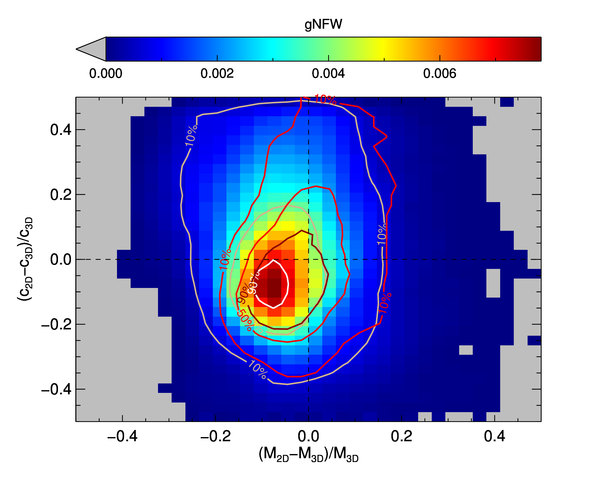}
   \includegraphics[width=0.33\hsize]{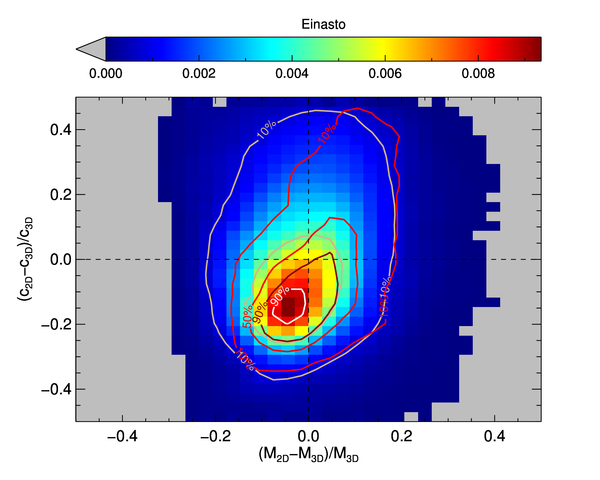}
 \caption{Distributions of {\tt MUSIC-2} halos in the plane $(c_{2D}-c_{3D})/c_{3D}$ vs $(M_{2D}-M_{3D})/M_{3D}$. Results are shown for the NFW, gNFW, and Einasto fits (left, central, and right panels, respectively). The two-dimensional histograms refer to the whole sample. The grey/white contours overlaid to the image show the intensity levels corresponding to $10\%$, $50\%$, and $90\%$ of the probability peak. The red contours correspond to the same levels for the distributions of the relaxed halos.} 
\label{fig:cm_shifts}
\end{figure*}

Potentially important differences between this work and \cite{2013ApJ...766...32B} are: 1)  our simulations include baryons, while the halos studied by \cite{2013ApJ...766...32B} are made only of dark-matter; 2) our analysis focuses on a limited mass range and the volume we sample is smaller compared to the simulations employed by \cite{2013ApJ...766...32B}; 3) the mass resolution of our simulations is roughly two orders of magnitude better; 4) \cite{2013ApJ...766...32B} fit their halos over a different radial range, $[0.1-1r_{vir}]$ vs. $[0.02-1r_{200}]$; and, finally, 5) \cite{2013ApJ...766...32B} fit the mass profiles instead of the density profiles, as we do. Given that  our simulations are non-radiative, it is unlikely that the differences between the $c-M$ relations  arise from baryonic effects. \cite{2013MNRAS.428.2921D} show that concentrations are higher by $5-15\%$ in radiative simulations compared to dark-matter only simulations. This result, however, was obtained using hydrodynamical simulations which are known to suffer of the over-cooling problem. It has been shown by other authors that halos in adiabatic simulations develop density profiles pretty similar to those of pure dark-matter halos \citep{2012MNRAS.427..533K}.  
The different mass range, volume and resolution of the simulations may have a larger impact on the results. Being our halos sampled with a larger number of particles, the profiles are better resolved. Thus, the measurements of the individual concentrations should be more robust, and allow us to resolve smaller radial scales. On the other hand, since their volume is bigger, \cite{2013ApJ...766...32B} have a larger number of massive halos to constrain the $c-M$ relation at the cluster scales.  On the contrary, \cite{2013ApJ...766...32B} fit  halos over 3 orders-of-magnitude in mass. It may be possible that the strong redshift evolution of their $c-M$ relation is driven by the smallest halos. Overall, it is likely that the higher normalisation of our $c-M$ is largely due to the better resolution of the {\tt MUSIC-2} sample compared to the  simulation sets used in     \cite{2013ApJ...766...32B}.
 
The bottom panel in Fig.~\ref{fig:3dvsBhat} shows another comparison between our best fit NFW $c-M-z$ relation and the results of  \cite{2013ApJ...766...32B}. The solid and the dotted lines show our relations for relaxed and the super-relaxed samples, respectively. The most striking difference with \cite{2013ApJ...766...32B} (dashed lines) is that we find a much stronger dependence of concentration on the halo dynamical state. While the normalisation of our $c-M-z$ relation increases by $\sim 10\%$ between the full and the relaxed samples, \cite{2013ApJ...766...32B} find that concentrations of relaxed halos increase only by $\sim 3\%$.  
 
\subsubsection{The concentration-mass relation in 2D} 
\label{sect:cm2d}

The 2D concentration-mass relation of the {\tt MUSIC-2} halos will be discussed in details in an upcoming paper (Vega et al., in prep). We briefly summarise  some  properties of this $c-M$ relation which are relevant for the following discussion. Projection effects do affect concentrations, which are generally found to be smaller than in 3D. This effect of triaxiality, discussed also in \cite{2012MNRAS.426.1558G}, is illustrated in Fig.~\ref{fig:cm_shifts}, where we show the distribution of the {\tt MUSIC-2} halos in the $(c_{2D}-c_{3D})/c_{3D}$ vs $(M_{2D}-M_{3D})/M_{3D}$ plane. The 2D histograms show that, regardless of the fitting model, the masses and concentrations derived from fitting the surface-density profiles tend to be smaller than measured from fitting the density profiles. The trend is in qualitative agreement with the findings of   \cite{2012MNRAS.426.1558G}, although the amplitude of both the concentration and mass biases found here is smaller.  The white contours overlaid to the 2D histograms show the intensity levels corresponding to $10\%$, $50\%$, and $90\%$ of the peaks of the distributions. The red contours indicate the same intensity levels for the subsample of super-relaxed halos. As we explained in Sect.~\ref{sect:masses}, the bias is reduced for the relaxed halos, because these systems are typically more spherical than the un-relaxed halos. 

The best fit parameters of the 2D $c-M-z$ relation are listed in  Table~\ref{tab:cmrelations}. The relations for halos at $z=0.250$ are given by the solid lines in Fig.~\ref{fig:cmfitssl}. Interestingly, the $c-M$ relation is very flat and characterised by an inverted slope compared to the $c-M$ relation in 3D. This suggests that the 2D concentrations underestimate the 3D ones more significantly at the lowest than at the highest masses.  One possible explanation is that the halo triaxiality is somehow biased  below the completeness limits listed in Table~\ref{table:masses}. However, we checked that the $c-M$ relation obtained only from halos above the completeness limits does not differ significantly from what we obtain using the extended sample. However, the constraints on its slope are obviously weakened. In addition, we notice that also \cite{2012MNRAS.426.1558G} find indications for a 2D concentration bias which decreases as a function of mass. This  will be discussed in Vega et al. (in prep.). 

\begin{deluxetable*}{ccccccc}[!t]
\tablecaption{Best fit parameters for the 3D and 2D $c-M-z$ relations. The results are listed  for the concentration-mass measurements based on the NFW, gNFW, and Einasto models. First column: fitting model; second column: 3D or 2D analysis; third column: relaxation state (all=full sample; rel=relaxed; srel=super-relaxed); columns 3,4,5: $c-M-z$ parameters (see Eq.~\ref{eq:cmfunction}); column 6: selection function (ext=extended sample, no selection applied except that based on the relaxation state; sl=strong lensing selection; xray=X-ray selection). The parameters of the 2D $c-M-z$ relation for the extended sample are taken from Vega et al. (in prep.).}
\tablehead{
\colhead{Fitting func.} & \colhead{3D/2D} & \colhead{relax.} & \colhead{$A$} & \colhead{$B$} & \colhead{$C$} & sel. func.\\ 
}
\startdata
NFW & 3D & all & $ 3.757 \pm  0.054 $ & $ 0.288 \pm  0.077$ & $-0.058\pm 0.017$ & ext\\
NFW & 3D & rel & $ 4.051 \pm  0.067 $ & $ 0.197 \pm  0.093$ & $-0.084\pm 0.020$ & ext\\
NFW & 3D & srel & $ 4.704 \pm  0.151 $ & $ 0.519 \pm  0.187$ & $-0.054\pm 0.039$ & ext\\
NFW & 2D & all & $ 3.580 \pm  0.040 $ & $ 0.003 \pm  0.053$ & $ 0.051\pm 0.013$ & ext\\
NFW & 2D & rel & $ 3.813 \pm  0.050 $ & $ 0.108 \pm  0.064$ & $-0.032\pm 0.015$ & ext\\
NFW & 2D & srel & $ 4.380 \pm  0.113 $ & $ 0.420 \pm  0.137$ & $-0.052\pm 0.030$ & ext\\
gNFW & 3D & all & $ 3.671 \pm  0.055 $ & $ 0.050 \pm  0.086$ & $ 0.101\pm 0.019$ & ext\\
gNFW & 3D & rel & $ 4.091 \pm  0.068 $ & $ 0.057 \pm  0.098$ & $ 0.018\pm 0.021$ & ext\\
gNFW & 3D & srel & $ 4.646 \pm  0.152 $ & $ 0.457 \pm  0.195$ & $-0.023\pm 0.040$ & ext\\
gNFW & 2D & all & $ 4.088 \pm  0.047 $ & $-0.228 \pm  0.055$ & $ 0.164\pm 0.014$ & ext\\
gNFW & 2D & rel & $ 4.261 \pm  0.055 $ & $-0.159 \pm  0.063$ & $ 0.071\pm 0.015$ & ext\\
gNFW & 2D & srel & $ 4.660 \pm  0.117 $ & $ 0.138 \pm  0.129$ & $ 0.022\pm 0.029$ & ext\\
Einasto & 3D & all & $ 3.407 \pm  0.055 $ & $ 0.040 \pm  0.092$ & $ 0.088\pm 0.020$ & ext\\
Einasto & 3D & rel & $ 3.805 \pm  0.068 $ & $ 0.088 \pm  0.104$ & $-0.007\pm 0.022$ & ext\\
Einasto & 3D & srel & $ 4.366 \pm  0.151 $ & $ 0.470 \pm  0.204$ & $-0.046\pm 0.043$ & ext\\
Einasto & 2D & all & $ 3.617 \pm  0.034 $ & $ 0.070 \pm  0.049$ & $ 0.103\pm 0.012$ & ext\\
Einasto & 2D & rel & $ 3.729 \pm  0.041 $ & $ 0.020 \pm  0.060$ & $ 0.028\pm 0.014$ & ext\\
Einasto & 2D & srel & $ 4.151 \pm  0.096 $ & $ 0.352 \pm  0.126$ & $ 0.012\pm 0.028$ & ext\\
NFW & 2D & all & $ 3.978 \pm  0.055 $ & $ 0.651 \pm  0.073$ & $-0.214\pm 0.018$ & sl\\
NFW & 2D & rel & $ 4.200 \pm  0.068 $ & $ 0.593 \pm  0.090$ & $-0.185\pm 0.021$ & sl\\
NFW & 2D & srel & $ 4.658 \pm  0.150 $ & $ 0.781 \pm  0.189$ & $-0.124\pm 0.041$ & sl\\
gNFW & 2D & all & $ 4.338 \pm  0.056 $ & $ 0.276 \pm  0.073$ & $-0.060\pm 0.018$ & sl\\
gNFW & 2D & rel & $ 4.571 \pm  0.069 $ & $ 0.310 \pm  0.089$ & $-0.053\pm 0.020$ & sl\\
gNFW & 2D & srel & $ 4.892 \pm  0.152 $ & $ 0.558 \pm  0.187$ & $-0.059\pm 0.041$ & sl\\
Einasto & 2D & all & $ 3.774 \pm  0.053 $ & $ 0.465 \pm  0.080$ & $-0.128\pm 0.019$ & sl\\
Einasto & 2D & rel & $ 3.961 \pm  0.066 $ & $ 0.489 \pm  0.098$ & $-0.128\pm 0.022$ & sl\\
Einasto & 2D & srel & $ 4.317 \pm  0.147 $ & $ 0.684 \pm  0.208$ & $-0.102\pm 0.045$ & sl\\
NFW & 2D & all & $ 4.105 \pm  0.100 $ & $ 0.668 \pm  0.341$ & $-0.160\pm 0.108$ & xray\\
gNFW & 2D & all & $ 4.228 \pm  0.138 $ & $ 0.376 \pm  0.458$ & $-0.080\pm 0.145$ & xray\\
Einasto & 2D & all & $ 3.880 \pm  0.119 $ & $-0.017 \pm  0.425$ & $-0.035\pm 0.137$ & xray\\
\enddata
\label{tab:cmrelations}
\end{deluxetable*}

\section{The concentration-mass relation of CLASH-like clusters}
We can now discuss how different cluster selection methods impact  the $c-M-z$ relation.  We will start with the $c-M-z$ relation of strong-lensing-galaxy clusters. Then, we investigate the $c-M-z$ relation obtained by selecting halos on the basis of their X-ray morphology. The results of this analysis are compared to the observations in \cite{2014arXiv1404.1376M} and \cite{2014arXiv1404.1375U}. 

\subsection{The $c-M-z$ relation of strong lensing halos}
As explained above, the CLASH cluster sample is composed by 25 galaxy clusters, of which only 5 were selected on the basis of their SL strength. The remaining 20 clusters are not SL selected and they were chosen on the basis of their X-ray morphology. We will discuss this selection method in the next section. Nevertheless, SL features (multiple images and arcs) have been securely detected in all CLASH clusters except RXJ1532.8+3021. The analysis of these SL features have allowed the creation of detailed lens models and the  measurement of their Einstein radii (Zitrin et al. 2014, in prep). The Einstein radii for sources at redshift $z_{\rm s}=2$ are within the range [5-55] arcsec.

We construct the $c-M-z$ relation of SL galaxy clusters by  selecting  those projections where we measure an Einstein radius compatible with those measured in the CLASH sample. As explained, the Einstein radius is defined as in Eq.~\ref{eq:einsteinr}. 

\begin{figure}[t]
\includegraphics[width=1\hsize]{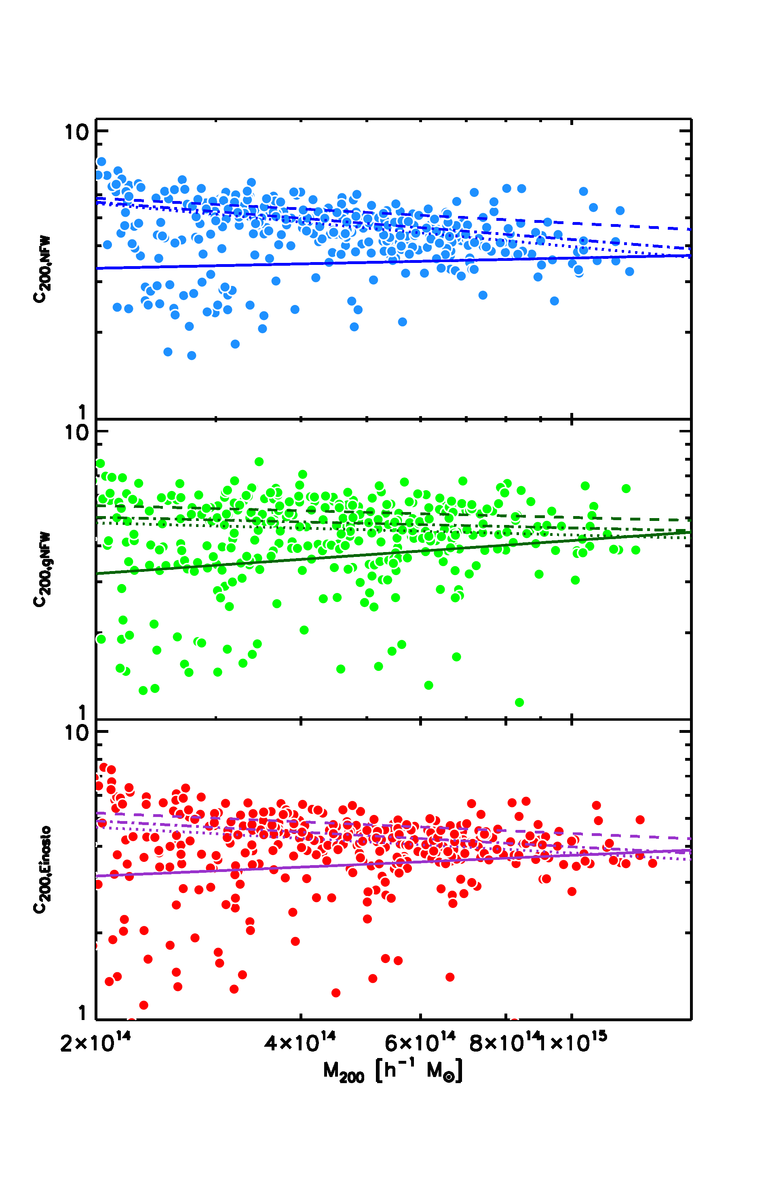}
\caption{Concentration mass relation for strong-lensing halos at $z=0.250$. The lines indicate the results obtained for the halos in the full, relaxed, and super-relaxed samples (dotted, dot-dashed and dashed lines, respectively). For comparison, the 2D $c-M$ relation derived for the full sample including non-strong lenses is shown by the solid line. The coloured circles correspond to the projections capable of producing critical lines for $z_s=2$. The upper, middle, and bottom panels refer to the NFW, gNFW, and Einasto fits.}
\label{fig:cmfitssl}
\end{figure}

In Fig.~\ref{fig:cmfitssl}, we show the concentration-mass relations at $z=0.250$ derived from SL halos in the {\tt MUSIC-2} sample. The relations are displayed for the NFW, gNFW, and Einasto models (upper, middle, and bottom panels). The corresponding parameters are listed in Table~\ref{tab:cmrelations}. The dotted, dashed, and dot-dashed lines indicate the relations obtained for the full, the relaxed, and the super-relaxed samples, respectively. For comparison, we show also the $c-M$ relation for the full sample, including also the non-strong lenses, and discussed in Sect.~\ref{sect:cm2d}. By imposing that the halos are strong lenses in their projections, we remove a large fraction of halos with low concentrations, obtaining relations characterised by a larger normalisation. In particular, an increasingly larger number of halos of small mass is unable to produce an appreciable SL signal. By removing them from the initial catalog, we restore the negative logarithmic slope of the $c-M$ relation. Due to this selection, the concentration scales with mass as $c\propto M^{-0.214\pm0.018}$. This result is in very good agreement with the theoretical predictions of \cite{2013arXiv1311.1205G} and \cite{2012MNRAS.420.3213O}, who estimated the lensing bias of the $c-M$ via semi-analytic calculations employing triaxial halos. For the gNFW and for the Einasto models, the concentration-mass relations are slightly flatter. 

Even for the SL halos, the normalisation of the $c-M-z$ relation depends on the relaxation state. The most relaxed system  have the largest concentrations. The differences between the $c-M-z$ relations of relaxed and un-relaxed halos are smaller than found earlier for the whole sample including non-strong lenses, though. This is due to the fact that in the SL selected sample the fraction of relaxed and super-relaxed halos is pretty high. At $z=0.250$, about $75\%$ of the SL projections belong to relaxed halos. The fraction of super-relaxed halos in this sample is $\sim27\%$. At $z=0.667$ the fractions of relaxed and super-relaxed halos are $\sim 60\%$ and $\sim13\%$, respectively. 


Finally, we find that the redshift evolution of the $c-M$ relation of SL halos is stronger than for non-SL halos. The values for the $B$ parameters listed in Table~\ref{tab:cmrelations} are in the range $[0.48-0.64]$ for the three fitting models.

\subsection{The $c-M-z$ relation of X-ray selected halos}
We discuss now the impact of the X-ray selection on the concentration-mass relation. In particular, we discuss the expectations for halos selected such to resemble the X-ray morphologies of the clusters in the CLASH X-ray selected sample. The results shown here are based on the analysis of three projections per halo, and the halos considered are those with 3D mass above the completeness limits given in Table~\ref{table:masses}. The restriction of the analysis to this smaller sample of simulated halos was dictated by the large computational time required to produce the X-ray simulated observations. The mass range covered by these simulations is however representative of the mass range of the CLASH clusters (see both Merten et al. 2014, and Umetsu et al. 2014).
 
\begin{figure}[t]
\includegraphics[width=1\hsize]{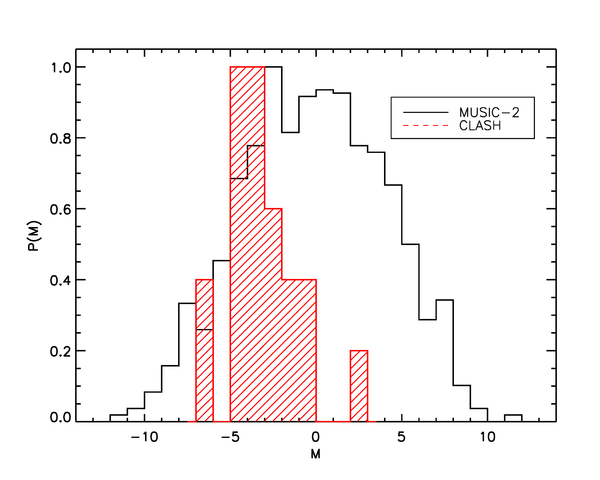}
\caption{Distributions of regularity parameters $M$ in the {\tt MUSIC-2} (black histogram) and in the CLASH sample (red histogram). }
\label{fig:reg_distr}
\end{figure}

\begin{figure}[t]
\includegraphics[width=1\hsize]{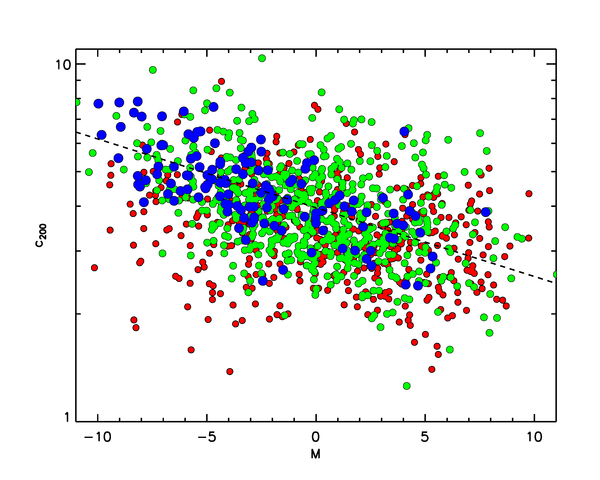}
\includegraphics[width=1\hsize]{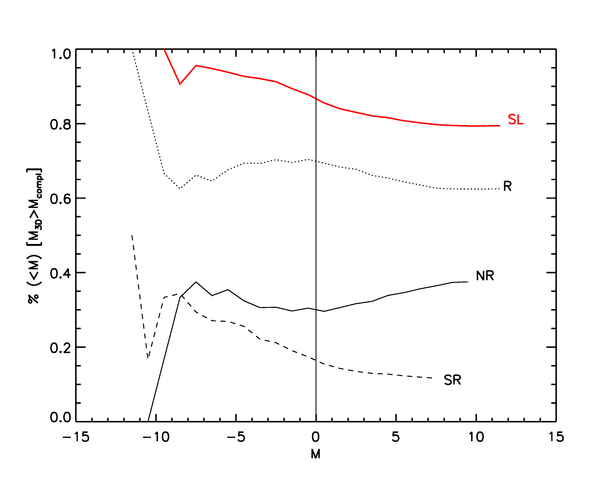}
\caption{Upper panel: correlation between projected concentration and regularity parameter M. The red, green, and blue circles indicate the un-relaxed, relaxed, and super-relaxed halos, respectively. The dashed line show the best linear fit  between $\log_{10}(c_{200})$ and $M$, obtained using of all the data points and given in Eq.~\ref{eq:corrm}. Bottom panel: fraction of strong lensing (SL), un-relaxed (NR), relaxed (R), and super-relaxed halos (SR) in samples selected by means of the $M$ parameter.}
\label{fig:cbias_mrel}
\end{figure}

As explained in Sect.~\ref{sect:xmorpho}, the X-ray morphology is measured by means of five morphological parameters. They can be combined to quantify the degree of regularity of the halos as shown in Eq.~\ref{eq:morphopar}.  
The regularity parameters of the CLASH clusters, as measured in their X-ray images, are listed in Table~\ref{tab:CLASH}. In Fig.~\ref{fig:reg_distr}, we show their distribution (red histogram) and we compare it to the distribution of the regularity parameters in the {\tt MUSIC-2} sample (black histogram). The histograms have been normalised to have the same peak value. As it emerges from these distributions, the CLASH clusters have quite typical regularity parameters  in the simulations. With the exception of MACSJ1206.2-0847, the clusters in the CLASH X-ray selected sample have negative $M$ parameters, indicating that they are more regular than the mean of the simulations. Even in the case of MACSJ1206.2-0847 other works based on different analyses find that this system is not likely to be perturbed by significant substructures \citep{2013ApJ...776...91L,2013A&A...558A...1B}.
This is expected given that these clusters were selected on the basis of their X-ray regularity. On the other hand, the comparison shows that their regularity is not extreme, in the sense that there are several simulated halos with regularity parameters exceeding the values for the CLASH clusters. In fact that the simulated sample has a tail of low $M$-values extending well beyond those of the CLASH clusters.

In the upper panel of Fig.~\ref{fig:cbias_mrel}, we show that the concentration inferred from the analysis of the 2D mass distributions is correlated with the regularity parameter $M$. The red, green, and blue circles refer to un-relaxed, relaxed, and super-relaxed halos. The correlation was evaluated by measuring the linear Pearson correlation coefficient  $P$ between the $\log_{10}c_{200,2D}$ and the $M$ values. It is stronger for the super-relaxed halos, for which we measure $P=-0.67$. For the relaxed and the full samples, we obtain $P=-0.46$ and $P=-0.39$, respectively. The best linear fit between the two parameters is
\begin{equation}
	\log_{10}c_{200,2D}=(0.598\pm0.009)-(0.019\pm0.002)\times M\;.
		\label{eq:corrm}
\end{equation}
If we refer to the average of all halos in the simulations ($M=0$ by construction), for negative values of $M$ we expect a positive concentration bias. Since the median value of the $M$ parameters of the CLASH clusters is $M_{\rm CLASH}=-3.44$, on the basis of Eq.~\ref{eq:corrm}, we can give an estimate of the expected concentration bias for the CLASH X-ray selected sample, which is 
\begin{equation}
\frac{c_{200,{\rm CLASH}}}{c_{200,2D}(M=0)}=1.11\pm0.03\;. 
\end{equation}

An interesting question is whether this concentration excess compared to the full sample arises from the selection of purely relaxed halos. The answer is already contained in  the upper panel of  Fig.~\ref{fig:cbias_mrel}: a selection based on the $M$ regularity parameter does not lead to the construction of  purely relaxed sample. Indeed, the left side of the diagram contains several red circles, indicating that un-relaxed halos can have $M<0$. The composition of samples selected by means of the $M$ parameter is shown in the bottom panel of Fig.~\ref{fig:cbias_mrel}. The curves show the fractions of relaxed (R), non-relaxed (NR) and super-relaxed (SR) halos in the samples with regularity parameter smaller than $M$. As indicated by the dotted and the solid black lines (R and NR halos), the fraction of relaxed and un-relaxed halos is nearly constant as a function of $M$. Thus, there is no  strong correlation between X-ray regularity and halo relaxation. In particular,  we find that only $\sim70\%$ of the halos among those with $M<0$ is relaxed\footnote[4]{We remind that relaxed halos are identified by means of the criteria described in Sect.~{\ref{sect:relaxation}}. By definition, super-relaxed halos are also included in this category.}. The remainder $\sim30\%$ of the halos are un-relaxed. As said, this composition is very similar that of the full sample. 

On the contrary, as indicated by the dashed line, the fraction of super-relaxed halos decreases as a function of $M$. Thus,   in samples of clusters selected to have regular X-ray morphologies, we expect to have a larger fraction of super-relaxed halos. Since these  typically have larger concentrations, we expect that the average concentration in a $M$-selected samples is higher than in the full sample. 

The red-solid line shows that also the fraction of strong lensing (SL) halos in $M$-selected samples decreases as function of $M$. This trend reflects the correlation between concentration and regularity parameter. Being the halos more concentrated, they more easily act as strong lenses. However, we notice that a correlation exists also between the concentrations and the $M$ parameters of the un-relaxed halos, although this is weaker than for the relaxed and super-relaxed halos. For un-relaxed halos the linear Pearson coefficient is $P=-0.22$, indicating that also among these halos, those with small $M$ parameter tend to have larger concentrations. In part, the classification of un-relaxed halos as regular is due to the different radial scales over which the relaxation and the regularity are evaluated. While the former is measured using all particles inside the virial radius, the second is meant to quantify the morphology of the cluster cores, within $500$ kpc. A fraction of halos with regular X-ray morphologies have significant sub-structures outside $500$ kpc, which implies that they are classified as un-relaxed. These substructures  explain the low concentrations of those un-relaxed halos which have small $M$. However, for $\sim42\%$ of the un-relaxed halos with $M<0$, we do not find evidence for sub-structures outside the region where we carry over the X-ray morphological classification. These halos generally have 2D concentrations higher than the average of the sample, indicating that the selection based on X-ray morphology may lead to the inclusion of un-relaxed objects which are elongated along the line-of-sight. Such a sample would then be affected by a small orientation bias.     

We use the $M$ parameters to create a sample of X-ray selected halos. These halos are drawn from the full {\tt MUSIC-2} sample such to reproduce the distribution of the $M$ parameters found for the CLASH clusters. In doing so, we take into account  the masses and redshifts of the CLASH clusters. The masses are  taken from Merten et al. (2014). A halo is selected if it has a suitable $M$ parameter and the mass inferred from the 2D analysis is within $3\sigma$ from the mass measured by CLASH. To account for the redshift distribution, given that the halos available for this analysis are simulated only at four redshifts, we create a match between each CLASH cluster and the nearest simulated redshift. The matches are listed in Table~\ref{tab:CLASH}.

As explained earlier in the paper, the X-ray analysis is limited to three orthogonal lines of sight per halo. Given that many more projections are available in the 2D analysis of the {\tt MUSIC-2} halos, we can improve our statistical power by increasing the number of projections used. To do so, we identify the projections whose lines of sights are within $20\deg$ from those selected in the X-ray analysis.

\begin{figure}[t]
\includegraphics[width=1\hsize]{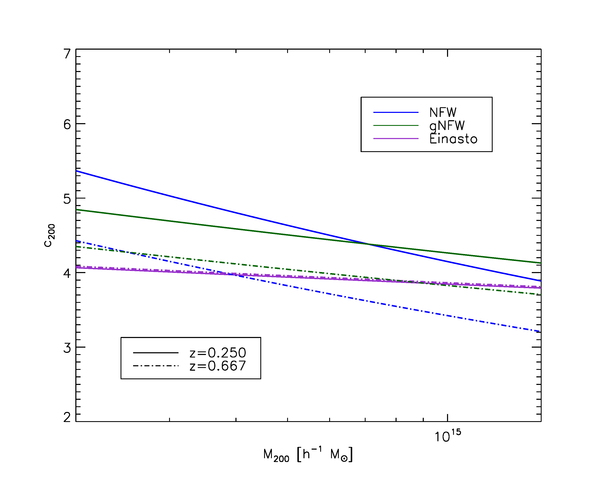}
\caption{Concentration mass relation at $z=0.250$  and $z=0.667$ for X-ray selected halos (solid and dot-dashed lines, respectively). The results are shown for the NFW, gNFW, and Einasto fitting models.}
\label{fig:cmfitsxray}
\end{figure}

Using the concentrations and the masses inferred from the 2D analysis of the X-ray selected projections, we fit the $c-M-z$ relation for our X-ray selected sample. The relation is shown in Fig.~\ref{fig:cmfitsxray} for all the fitting models employed in this study. The best fit parameters are listed in Table~\ref{tab:cmrelations}.  Overall, the $c-M-z$ relation for X-ray selected halos is in good agreement with the SL $c-M-z$ relation for a sample composed by both relaxed and un-relaxed halos. This is not surprising given that  X-ray selected halos are frequently efficient strong lenses, with  only $\sim8\%$ of them which do not have an extended critical line for sources at $z=2$. About  $70\%$ of the selected projections are  belonging to relaxed halos. About $18\%$ of them correspond to halos classified as super-relaxed.  
For the NFW model, we find that the concentrations scale with mass as $c\propto M^{-0.16\pm0.11}$, resulting in average concentrations which are intermediate between those predicted in 3D for relaxed and super-relaxed halos in the mass range  $2\times10^{14}\lesssim M_{200}\lesssim 10^{15}h^{-1}M_{\odot}$. 

Some  differences between the fitting models are found with regards to the redshift evolution of the concentration-mass relation. For the NFW model, the $c-M-z$ relation is evolving strongly. The redshift dependence is shallower in the case of the gNFW model, and it is consistent with zero evolution for the Einasto profile.  

\begin{deluxetable*}{lccccc}[!t]
\tablecaption{Comparison between CLASH clusters and X-ray selected halos. Column 1: cluster name; column 2: reference redshift in the simulations; column 3: true redshift of the CLASH cluster; column 4: Regularity parameter $M$; column 5: mass range of X-ray selected clusters in the simulation; column 6: mean NFW concentration of selected halos.}
\tablehead{
\colhead{Cluster } & \colhead{$z_{\rm sim}$} & \colhead{$z$} & \colhead{$M$} & \colhead{$M_{200,X}$} & \colhead{$c_{200,X}$} \\ 
 &  &  &  & \colhead{$[10^{14}\;h^{-1}M_{\odot}]$} &  \\ 
}
\startdata
      Abell383 & 0.250 & 0.188 &    -6.49 & $    8.52\pm    1.47$ & $    3.46\pm    1.09$\\
      Abell209 & 0.250 & 0.206 &    -0.87 & $    9.43\pm    1.76$ & $    4.09\pm    0.94$\\
     Abell1423 & 0.250 & 0.213 &    -3.11 & $    7.00\pm    1.80$ & $    4.60\pm    1.12$\\
     Abell2261 & 0.250 & 0.225 &    -3.93 & $    9.98\pm    2.03$ & $    3.76\pm    1.00$\\
  RXJ2129+0005 & 0.250 & 0.234 &    -3.70 & $    6.12\pm    2.71$ & $    3.69\pm    1.01$\\
      Abell611 & 0.250 & 0.288 &    -4.27 & $    8.50\pm    1.59$ & $    3.12\pm    1.43$\\
   MS2137-2353 & 0.333 & 0.313 &    -5.00 & $   10.41\pm    2.65$ & $    4.38\pm    1.11$\\
RXJ1532.8+3021 & 0.333 & 0.345 &    -6.27 & $    6.19\pm    2.65$ & $    3.73\pm    1.11$\\
 RXCJ2248-4431 & 0.333 & 0.348 &    -1.56 & $   11.50\pm    3.33$ & $    3.62\pm    1.09$\\
MACSJ1115+0129 & 0.333 & 0.352 &    -2.87 & $    9.00\pm    1.80$ & $    3.07\pm    1.45$\\
  MACSJ1931-26 & 0.333 & 0.352 &    -4.37 & $    6.92\pm    2.31$ & $    3.91\pm    1.05$\\
MACSJ1720+3536 & 0.429 & 0.391 &    -4.12 & $    7.50\pm    1.92$ & $    5.68\pm    1.81$\\
  MACSJ0429-02 & 0.429 & 0.399 &    -3.50 & $    8.05\pm    1.81$ & $    3.74\pm    1.10$\\
  MACSJ1206-08 & 0.429 & 0.439 &     2.29 & $    8.62\pm    1.96$ & $    3.14\pm    1.43$\\
  MACSJ0329-02 & 0.429 & 0.450 &    -2.90 & $    7.31\pm    1.89$ & $    3.82\pm    1.09$\\
  RXJ1347-1145 & 0.429 & 0.451 &    -2.79 & $   11.47\pm    4.20$ & $    3.62\pm    1.16$\\
  MACSJ1311-03 & 0.429 & 0.494 &    -3.44 & $    6.09\pm    2.31$ & $    3.90\pm    1.02$\\
  MACSJ1423+24 & 0.667 & 0.545 &    -4.10 & $    5.71\pm    2.49$ & $    3.93\pm    1.07$\\
  MACSJ0744+39 & 0.667 & 0.686 &    -1.56 & $    7.00\pm    1.93$ & $    4.58\pm    1.22$\\
\enddata
\label{tab:CLASH}
\end{deluxetable*}

\section{Predictions for individual {\tt CLASH} clusters}
Finally, we use the {\tt MUSIC-2} halos and their X-ray morphology to predict the concentrations of each individual CLASH cluster. As explained in Sect.~\ref{sect:xmorpho}, this is done using the parameter $C_X$, which measures the distance of each simulated halo from a given CLASH cluster in the multi-dimensional space defined by the X-ray morphological parameters. We select projections with $C_X<0.4$ to create the match. 

\begin{figure}[t]
\includegraphics[width=1\hsize]{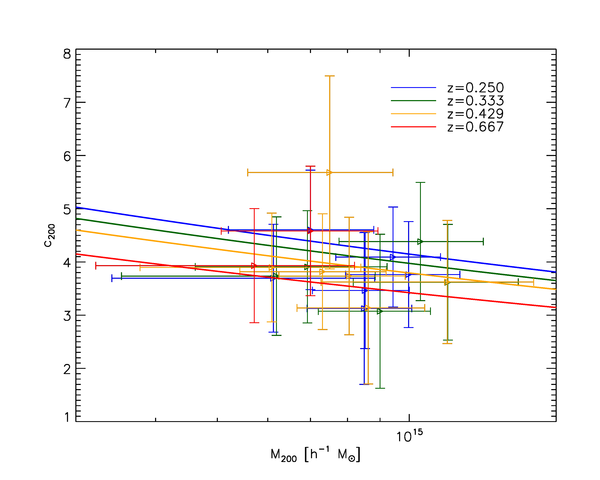}
\caption{NFW concentrations and masses of {\tt MUSIC-2} halos matching the X-ray morphologies of the CLASH X-ray selected clusters. The error bars reflect the scatter in the masses and concentrations of the halos matching each CLASH cluster. The data points have different colors depending on the redshift of the simulations. For comparison, we also show the $c-M-z$ relation derived from the simulated X-ray selected sample, whose parameters are given in Table~\ref{tab:cmrelations} (solid lines).}
\label{fig:cmclash}
\end{figure} 

Again, for each of the matched X-ray images, we include in our analysis the projections from nearby lines-of-sight. To be associated with a specific CLASH cluster, the halos must also have compatible masses and redshifts. For all the CLASH clusters except CLJ1226+3332, we could create associations with $\sim10-200$ projections.  CLJ1226+3332 turned out not to have any counterpart in the simulated set. For this cluster, Merten et al. (2014) measured a large mass, $M_{200}\sim 1.5\times10^{15}$. The cluster is also at high redshift ($z=0.89$) and due to the limited volume of the {\tt MultiDark} cosmological box, there are too few massive systems at such large redshift to make a fair comparison based on the X-ray morphology.

Having built the associations between simulated and real clusters, we estimate the concentrations by averaging over the selected projections. The results are listed in the 6$th$ column of Table~\ref{tab:CLASH} for the NFW model. In the $5th$ column, we report the mass range of the selected halos. On the basis of these results, we find that CLASH-like clusters have concentrations in the range $\sim [3-6]$. 
These measurements are shown in Fig.~\ref{fig:cmclash}. The different colours  allow to discriminate between the redshifts of the simulations. For comparison, we also show the $c-M-z$ relation previously determined using the larger sample of X-ray selected halos. 

The X-ray morphology may reflect  the orientation of the cluster. Clusters may appear to have round X-ray iso-photes if they have prolate three-dimensional shapes and have they major axis aligned with the line of sight. Knowing the shapes and the orientation of the {\tt MUSIC-2} halos we can estimate if a sample constructed to resemble the morphology of the CLASH clusters is likely to be affected by a large orientation bias. On the basis of the associations we made between real and simulated cluster, we find that the mean angle between the major axes of the simulated halos and the line of sight is $\sim54\deg$. This indicates that the orientation bias is modest. 

\section{Summary and conclusions}

In this paper, we used  a large set of 1419 cluster-sized halos evolved in N-body/hydrodynamical simulations and distributed over the redshift range $0.25\leq z \leq 0.67$, to make predictions about several properties of the clusters included in the CLASH sample \citep{2012ApJS..199...25P}. The simulations used here, which are taken from the  {\tt MUSIC-2} sample \citep{2013MNRAS.429..323S}, intentionally do not include radiative physics to avoid an artificial boost of the halo concentrations due to the well known over-cooling problem. 

First, we characterised the halos by studying their total density profiles. We fitted the profiles using three fitting models: the NFW, the gNFW, and the Einasto profiles. We derived concentration-mass relations and we quantified their dependence on the degree of relaxation. By fitting with the gNFW and with the Einasto profiles, we could also investigate the distribution of the inner slopes and of the shape parameter of the density profiles.  

We combined our work with measurements of concentrations and masses taken from Vega et al. (in prep.). These measurements were obtained by fitting the surface-density profiles extracted from  hundreds of projections of the {\tt MUSIC-2} halos. The fits were performed with  the same codes used to measure the surface-density profiles recovered from the strong and weak lensing analyses of the CLASH cluster sample, as described in Merten et al. (2014). The radial ranges over which the fits were performed are compatible with those used in the observational analysis. 

Using the X-MAS code \citep{GA04.1,2011ApJ...729...45R}, we produced simulated {\em Chandra} observations for three orthogonal lines-of-sight to  each  halo above the {\tt MUSIC-2} mass completeness limit. These simulated observations were processed using the same routines employed in Donahue et al. (in prep.) to carry out the X-ray morphological analysis of the CLASH clusters. The X-ray morphology of the simulated halos was quantified by means of five morphological parameters, which we combined to define a global regularity parameter. 

Using the concentrations and masses derived from the analysis of the surface-density profiles, we derived lensing-like concentration-mass relations including the effects of selection functions aimed at reproducing some observational properties of the CLASH clusters. In particular, we focused on their ability to produce strong lensing effects and  their X-ray regularity. For this purpose, we created two sub-samples of {\tt MUSIC-2} halos. The first includes halos with Einstein radii in the range of those of the CLASH clusters. The second is constructed such to reproduce the distribution of X-ray regularity parameters of the CLASH clusters.

Our results can be summarised as follows:
\begin{itemize}
\item We find that a large fraction of {\tt MUSIC-2} halos has density profiles which are better fitted by gNFW and Einasto profiles than by NFW profiles. Not surprisingly, the halos which mostly deviate from the NFW model are the least relaxed. For these halos more flexible profiles are needed to better reproduce the shape of the density profiles. The analysis based of the gNFW model shows that the inner slopes of these profiles are distributed over a wide range of values. On average, the  logarithmic inner slope is largely consistent with the NFW slope, though. The Einasto profile fits the halos slightly better than the gNFW model;
\item when seen in projection, the distribution of the inner slopes widens further, and a large fraction of halos is fitted with profiles that are flatter than the NFW at small radii. On average, the inner logarithmic slopes derived from the gNFW fits of the surface-density profiles is $\sim15\%$ smaller than found fitting the density profiles. About $15\%$ of the halos have inner logarithmic slopes smaller than 0.5;
\item the masses derived from the fits of the density profiles match quite well the true masses of the halos, with a scatter which is of only few percent. When they are recovered from the projected mass distributions, mimicking the results obtainable from the analysis of surface-density fields reconstructed via lensing, the masses are smaller than the true masses by less than $5\%$ on average. As discussed in \cite{2012MNRAS.426.1558G} a mass bias is expected for randomly oriented prolated triaxial halos. However, the amplitude of the bias for this sample is $\sim 50\%$ smaller than expected from semi-analytical calculations. The bias is even smaller for relaxed halos, because their shapes are more spherical;
\item the concentrations derived from the fits of the density profiles with different models are rather similar. However, we find that Einasto concentrations are smaller by $10-15\%$ compared to the NFW and gNFW concentrations;      
\item we find that the {\tt MUSIC-2} halos follow an intrinsic concentration-mass relation characterised by a slightly larger normalisation  compared to other concentration-mass relations recently proposed in the literature for the NFW model. The redshift evolution is rather weak.
\item  when we mimic the selection of clusters on the basis of their strong lensing signal,  we find that the concentration-mass relation derived from the analysis of the projected mass distributions is considerably steeper than expected for non-strong lenses. It also has a larger normalisation. This result holds for all the fitting models used in this work;
\item  using the X-ray regularity parameter $M$ to select halos with regular X-ray morphologies leads to the inclusion of both relaxed and un-relaxed halos in the sample. Therefore, the X-ray morphology, especially if evaluated in a relatively small region around the cluster centre, is not ideal at identifying relaxed halos; 
\item the parameter $M$ is correlated to the halo 2D concentration. The most regular halos have higher mass concentrations compared to the full sample of simulated halos, as they could be measured from a lensing analysis. The excess of concentration is explained in terms of 1) the higher fraction of super-relaxed objects in the X-ray selected sample and 2) to the presence, among the selected halos, of un-relaxed systems which happen to be well aligned with the line-of-sight. For a regularity parameter $M$ equal to the median value measured for the CLASH sample, we expect that the concentration will be higher than the average of all halos in the simulated set by $\sim 11\pm3\%$;
\item measuring the concentration-mass relation and its redshift evolution in a sub-sample of {\tt MUSIC-2} halos which reproduces the distribution of  X-ray regularity parameters of the clusters in the CLASH X-ray selected sample, we find that this has an amplitude and  mass dependence similar to those of  the concentration-mass relation of strong-lensing clusters.  We verified that the sample of X-ray selected halos is largely composed by strong lensing clusters, and contains a fraction of only $8\%$ of halos which do not have extended critical lines for sources at $z\sim2$.
\item the sample of X-ray selected halos is in large fraction composed by relaxed halos. These amount to $\sim70\%$ of sample.   
\end{itemize}

These results suggest that the CLASH clusters are prevalently relaxed and likely to be modestly affected by strong lensing bias. Once accounted for projection and selection effects, their NFW concentrations are expected to scale with mass as $c\propto M^{-0.16\pm0.11}$ for the NFW model, resulting in average concentrations which are intermediate between those predicted in 3D for relaxed and super-relaxed halos in the mass range  $2\times10^{14}\lesssim M_{200}\lesssim 10^{15}h^{-1}M_{\odot}$. Matching the simulations to the individual CLASH clusters on the basis of the X-ray morphology, we expect that the NFW concentrations recovered from the lensing analysis of the CLASH clusters are in the range $[3-6]$, with an average value of $3.87$ and a standard deviation of $0.61$. The median value of the concentrations in the simulated sample is $3.76$ and the first and third quartiles of the concentration distribution are 3.62 and 3.93, respectively. As shown in \cite{2010A&A...519A..90M} and in \cite{2008AJ....135..664H}, strong lensing  clusters are expected to be frequently elongated along the line of sight. For the simulated CLASH sample, the median angle between the major axis of the halos and the lines of sight selected from the X-ray analysis is $54 \deg$. This indicates that the orientation bias is very modest. It is consistent with the results based on the analysis of the halos from the {\tt MareNostrum Universe} presented in \cite{2010A&A...519A..90M}.  

\acknowledgments

The research was in part carried out at the Jet Propulsion Laboratory,
California Institute of Technology, under a contract with the National Aeronautics and Space Administration. MM thanks ORAU and NASA for supporting his research at JPL. MM, CG, LM acknowledge support from the contract ASI/INAF I/023/12/0, INFN/PD51, and the PRIN MIUR 2010Ð2011 ÔThe dark Universe and the cosmic evolution of baryons: from current surveys to EuclidÕ. ER acknowledges support from the National Science Foundation AST-1210973, SAO TM3-14008X (issued under NASA Contract No. NAS8-03060). CG's research is part of the project GLENCO, funded under the European Seventh Framework Programme, Ideas, Grant Agreement n. 259349. K.U. acknowledges support from the National Science Council of Taiwan (grant NSC100-2112-M-001-008-MY3) and from the Academia Sinica Career Development Award.
Support for AZ is provided by NASA through Hubble Fellowship grant \#HST-HF-51334.01-A awarded by STScI.
D.G., S.S. and P.R. were supported by SFBTransregio 33 'The Dark Universe' by the Deutsche Forschungsgemeinschaft (DFG) 
and the DFG cluster of excellence 'Origin and Structure of the Universe'.
This work was supported in part by contract research ``Internationale
Spitzenforschung II/2-6'' of the Baden W\"{u}rttemberg Stiftung.
The Dark Cosmology Centre is funded by the DNRF. JS was supported by NSF/AST1313447, NASA/NNX11AB07G, and the Norris Foundation CCAT Postdoctoral Fellowship. The  MUSIC simulations were performed at the Barcelona Supercomputing Center (BSC) and the initial conditions were done at the Leibniz Rechenzentrum Munich (LRZ).  GY  and FS acknowledge  support from MINECO under research grants  AYA2012-31101,  FPA2012-34694 and MultiDark CSD2009-00064. We thank Stefano Borgani and the whole computational astrophysics group at the University of Trieste and at INAF-OATS for giving us access to their set of hydrodynamical simulations.

\bibliographystyle{apj_fewer_names}
\bibliography{/Users/maxmen/Documents/Pub/Papers/TeXMacro/master}

\end{document}